\documentclass[sigconf]{acmart} 

\usepackage{graphicx}
\usepackage{balance}  
\usepackage{graphics}
\usepackage[ruled, vlined, linesnumbered]{algorithm2e} 
\usepackage{soul}
\usepackage{float}
\usepackage{amsmath}
\usepackage{amsfonts}
\usepackage{amsbsy}
\usepackage{amsthm}
\usepackage{multirow}
\usepackage[mathscr]{eucal} 

\usepackage{xspace}
\usepackage{booktabs}
\usepackage{subcaption}
\usepackage[]{caption}
\usepackage{mathtools}
\usepackage{tabularx}
\usepackage{verbatim}
\usepackage{enumitem}
\usepackage{array}
\usepackage{dsfont}
\usepackage{url}

\AtBeginDocument{%
  \providecommand\BibTeX{{%
    \normalfont B\kern-0.5em{\scshape i\kern-0.25em b}\kern-0.8em\TeX}}}

\copyrightyear{2023}
\acmYear{2023}
\setcopyright{acmlicensed}\acmConference[KDD '23]{Proceedings of the 29th ACM SIGKDD Conference on Knowledge Discovery and Data Mining}{August 6--10, 2023}{Long Beach, CA, USA}
\acmBooktitle{Proceedings of the 29th ACM SIGKDD Conference on Knowledge Discovery and Data Mining (KDD '23), August 6--10, 2023, Long Beach, CA, USA}
\acmPrice{15.00}
\acmDOI{10.1145/3580305.3599274}
\acmISBN{979-8-4007-0103-0/23/08}

\DeclareMathOperator*{\argmax}{arg\,max}

\newcommand{\cat}[0]{\mathbin\Vert}



\setcopyright{none}
\renewcommand\footnotetextcopyrightpermission[1]{} 
\begin{document}

    \title{Classification of Edge-dependent Labels of Nodes in Hypergraphs}
    
    \settopmatter{authorsperrow=4}
    \author{Minyoung Choe}
	\affiliation{%
		\institution{KAIST}
		\city{Seoul}
		\country{Korea}
	}
	\email{minyoung.choe@kaist.ac.kr}
	
	\author{Sunwoo Kim}
	\affiliation{%
    	\institution{KAIST}
		\city{Seoul}
		\country{Korea}
	}
	\email{kswoo97@kaist.ac.kr}
	
	\author{Jaemin Yoo}
	\affiliation{%
		\institution{Carnegie Mellon University}
		\city{Pittsburgh, PA}
		\country{USA}
	}
	\email{jaeminyoo@cmu.edu}
	
	\author{Kijung Shin}
	\affiliation{%
		\institution{KAIST}
		\city{Seoul}
		\country{Korea}
	}
	\email{kijungs@kaist.ac.kr}
	
    \begin{abstract}
		A hypergraph is a data structure composed of nodes and hyperedges, where each hyperedge is an any-sized subset of nodes.
Due to the flexibility in hyperedge size, hypergraphs represent group interactions (e.g., co-authorship by more than two authors) more naturally and accurately than ordinary graphs.
Interestingly, many real-world systems modeled as hypergraphs contain edge-dependent node labels, i.e., node labels that vary depending on hyperedges.
For example, on co-authorship datasets, 
the same author (i.e., a node) can be the primary author in a paper (i.e.,  a hyperedge) but the corresponding author in another paper (i.e.,  another hyperedge).

In this work, we introduce a classification of edge-dependent node labels as a new problem. This problem can be used as a benchmark task for hypergraph neural networks, which recently have attracted great attention,
and also the usefulness of edge-dependent node labels has been verified in various applications. 
To tackle this problem, we propose \method, a novel hypergraph neural network
that represents the same node differently depending on the hyperedges it participates in by reflecting its varying importance in the hyperedges. To this end, \method models the relations between nodes within each hyperedge, 
using their relative centrality as positional encodings.
In our experiments, we demonstrate that \method significantly and consistently outperforms ten competitors on six real-world hypergraphs, and
we also show successful applications of
\method to (a) ranking aggregation, (b) node clustering, and (c) product return prediction.



	\end{abstract}
	
	\newcommand\red[1]{\textcolor{red}{#1}}
\newcommand\orange[1]{\textcolor{orange}{#1}}
\newcommand\blue[1]{\textcolor{blue}{#1}}
\newcommand\green[1]{\textcolor{green}{#1}}
\newcommand\gray[1]{\textcolor{gray}{#1}}
\newcommand\kijung[1]{\textcolor{red}{[Kijung: #1]}}
\newcommand\minyoung[1]{\textcolor{blue}{#1}}
\newcommand\geon[1]{\textcolor{brown}{#1}}
\newcommand\jaemin[1]{\textcolor{orange}{[Jaemin: #1]}}

\newcommand{\smallsection}[1]{{\vspace{0.02in} \noindent {{\underline{\smash{\bf #1.}}}}}}
\newtheorem{obs}{\textbf{Observation}}
\newtheorem{defn}{\textbf{Definition}}
\newtheorem{thm}{\textbf{Theorem}}
\newtheorem{axm}{\textbf{Axiom}}
\newtheorem{lma}{\textbf{Lemma}}
\newtheorem{cor}{\textbf{Corollary}}
\newtheorem{problem}{\textbf{Problem}}
\newtheorem{pro}{\textbf{Problem}}

\newcommand\und[1]{\underline{#1}}

\newcommand{\SM}{\mathcal{M}}%
\newcommand{\SG}{\mathcal{G}}%
\newcommand{\SSS}{\mathcal{S}}%
\newcommand{\SV}{\mathcal{V}}%
\newcommand{\SE}{\mathcal{E}}%
\newcommand{\SL}{\mathcal{L}}%
\newcommand{\SGH}{\mathcal{\hat{G}}}%
\newcommand{\SEH}{\mathcal{\hat{E}}}%
\newcommand{\SVH}{\mathcal{\hat{V}}}%
\newcommand{\SEVH}{\SE({\SVH})}%
\newcommand{\SVEH}{\SV({\SEH})}%
\newcommand{\SD}{\mathcal{D}}%

\newcommand{\fhat}{\hat{f}}%
\newcommand{\flargehat}{\hat{F}}%
\newcommand{\yhat}{\hat{y}}%

\newcommand{\Ghat}{\SGH=(\SVH,\SEH)}%
\newcommand{\Glong}{\SG=(\SV,\SE)}%

\newcommand{\cmark}{\ding{51}}%
\newcommand{\xmark}{\ding{55}}%

\newcommand{\method}{\textsc{WHATsNet}\xspace}
\newcommand{\attention}{\textsc{ATT}\xspace}
\newcommand{\aggregation}{\textsc{AGG}\xspace}
\newcommand{\MultiheadAtt}{\textsc{MAB}\xspace}
\newcommand{\WithinAttentionFull}{within attention\xspace}
\newcommand{\WithinAttention}{\textsc{WithinATT}\xspace}
\newcommand{\PE}{\textsc{WithinOrderPE}\xspace}

\newcommand{\methodgrid}{\textsc{MiDaS-Grid}\xspace}
\newcommand{\methodauto}{\textsc{MiDaS}\xspace}

\definecolor{myred}{RGB}{195, 79, 82}
\definecolor{mygreen}{RGB}{86, 167 104}
\definecolor{myblue}{RGB}{74, 113 175}

\newcommand{\bigcell}[2]{\begin{tabular}{@{}#1@{}}#2\end{tabular}}

\let\oldnl\nl
\newcommand{\nonl}{\renewcommand{\nl}{\let\nl\oldnl}}

	\maketitle
    
    \section{Introduction}
	\label{sec:intro}
\begin{figure}[t]
    \begin{minipage}{0.63\linewidth}
        \includegraphics[width=1.0\textwidth]{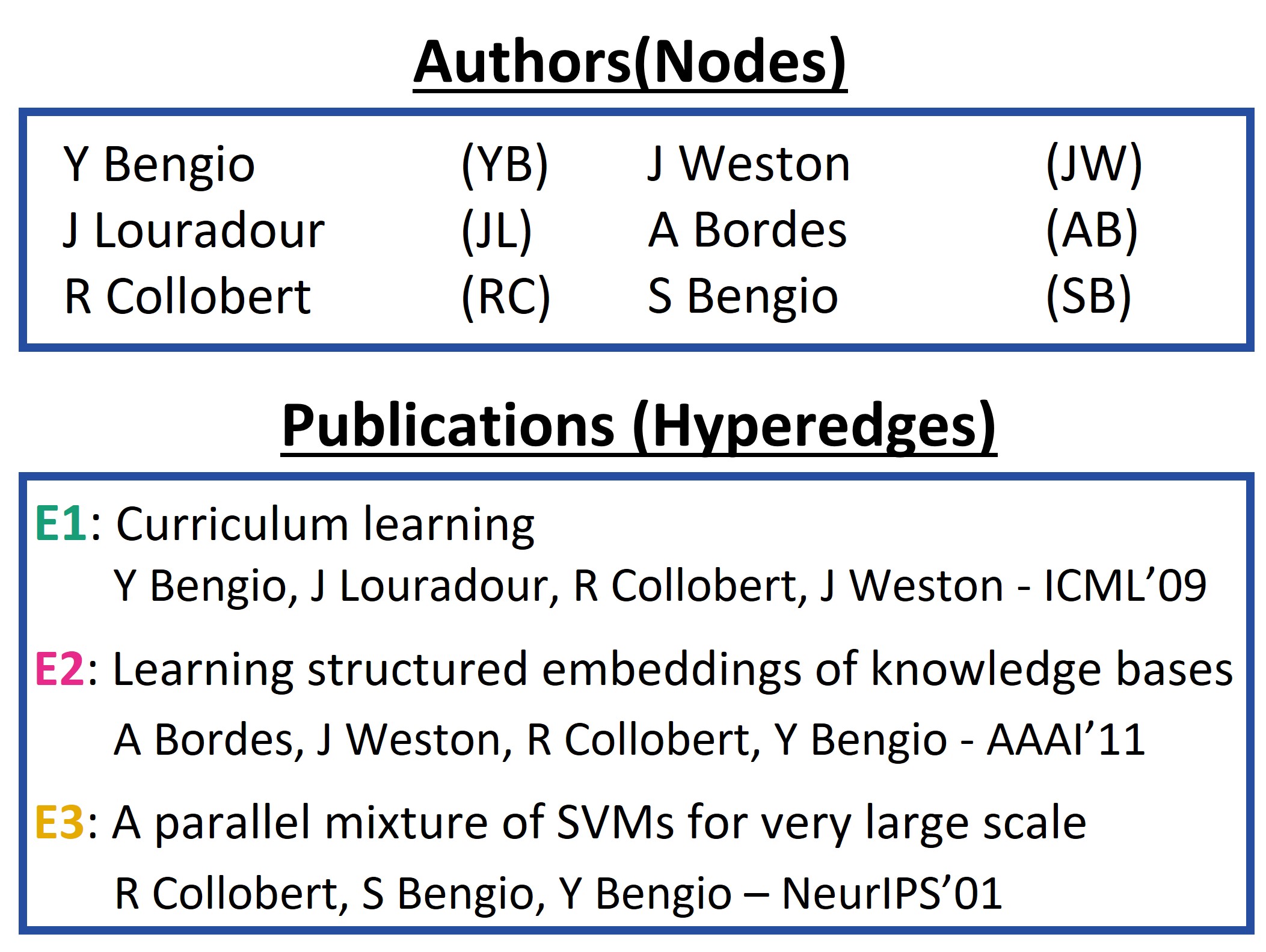}
    \end{minipage}
    \begin{minipage}{0.34\linewidth}
        \vspace{0.2\linewidth}
        \includegraphics[width=1.0\textwidth]{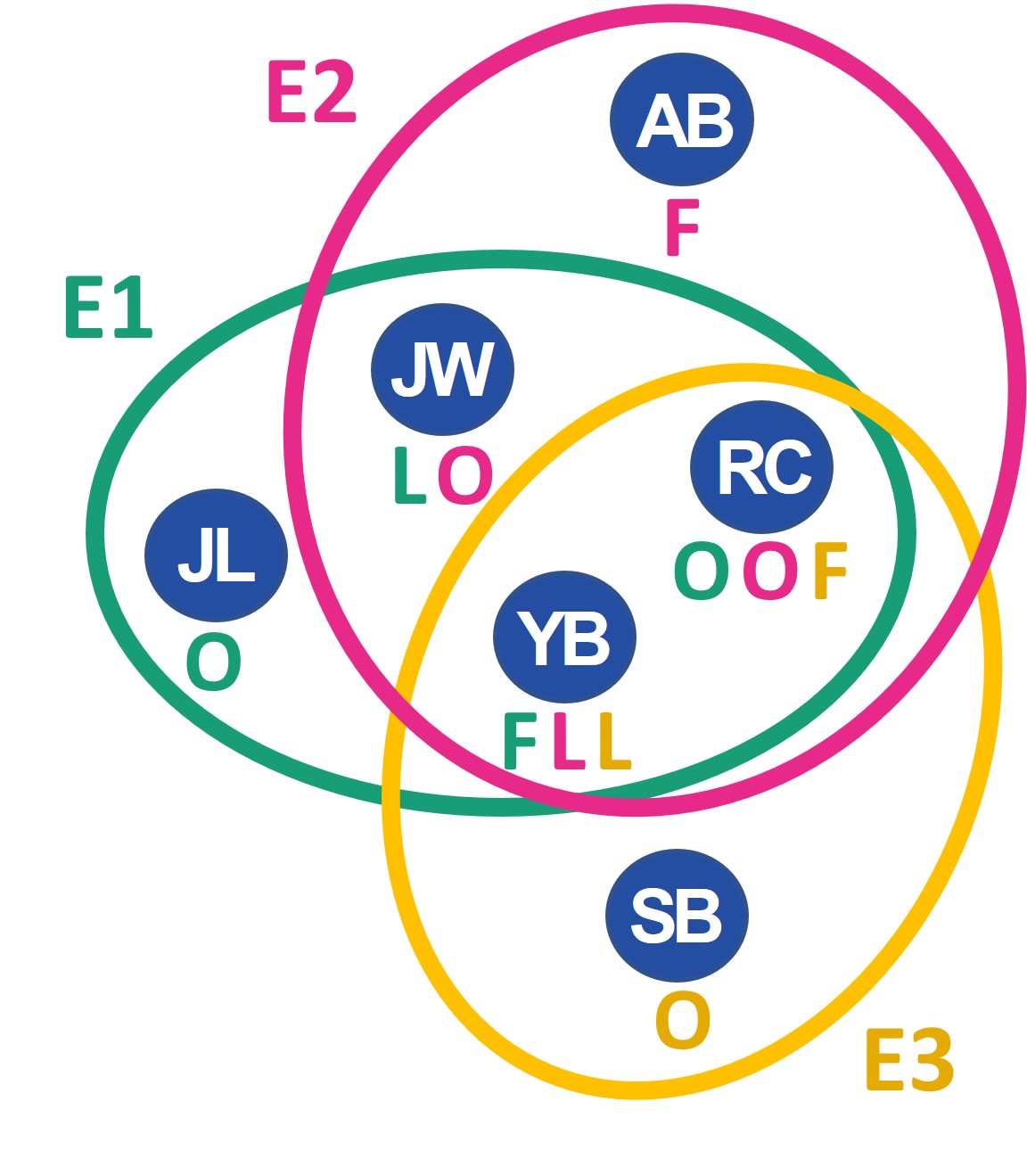}
    \end{minipage}
    
    \begin{minipage}{1.3\linewidth} \small
         \hspace{0.02\linewidth} (a) Example data: Coauthors in publications \hspace{0.06\linewidth} (b) Hypergraph
    \end{minipage}
    
    \caption{
    The three publications by the six authors in (a) are modeled as the hypergraph with six nodes and three hyperedges in (b). The labels of nodes indicate the orders of the authors (\underline{F}irst, \underline{L}ast, or \underline{O}thers) in each publication, and thus they are dependent on hyperedges.
    }
    \label{fig:intro_ex}
\end{figure}

Real-world relationships are complex and often go beyond pairwise relations.
For example, a research paper is usually coauthored by a group of researchers, and an email is often sent to multiple receivers.
A \textit{hypergraph} is a natural representation of such group relations \cite{benson2018sequences,benson2018simplicial,lee2021hyperedges,do2020structural,lee2020hypergraph}.
It consists of nodes and hyperedges, and each hyperedge is a set of any number of nodes (see Figure~\ref{fig:intro_ex}).
Hypergraph modeling is used in various fields, including recommendation systems~\cite{xia2021self,hada2021hypertenet,qiu2021exploiting}, and physics~\cite{gu2020quantum}, leading to
better performance than
ordinary-graph modeling in node clustering~\cite{wolf2016advantages}, interaction prediction~\cite{yoon2020much}, anomaly detection \cite{lee2022hashnwalk}, etc.

To leverage the advantages of hypergraph modeling,
several hypergraph neural networks \cite{feng2019hypergraph,yadati2019hypergcn,dong2020hnhn,bai2021hypergraph,huang2021unignn,chien2021you,hwang2021hyfer,aponte2022hypergraph} have been proposed.
They commonly involve propagating node embeddings to incident hyperedges for updating, which are subsequently propagated back to the nodes.
Most of them have been evaluated on node classification tasks, focusing on the global properties of nodes.

However, in many real-world hypergraphs, node properties 
vary depending on the hyperedges they are involved in.
For instance, in co-authorship, the same researcher can be the primary author in one paper but the corresponding author in another paper, as shown in Figure~\ref{fig:intro_ex}. 
In emails, the same person can be a receiver or a sender in different emails, and in online Q\&A platforms, a person can be a questioner and an answerer in different posts.
These edge-dependent node properties have proven useful in various tasks,
such as ranking aggregation~\cite{chitra2019random}, node clustering~\cite{hayashi2020hypergraph}, product-return prediction~\cite{li2018tail}, and anomaly detection~\cite{lee2022hashnwalk}.

In this work, we introduce a classification of edge-dependent node labels 
as a new problem for hypergraphs.
The problem works as an effective hypergraph-learning benchmark task that is complementary to common benchmark tasks (e.g., node classification, hyperedge prediction, and clustering) for three reasons.
First, it evaluates the capability of models in capturing features unique to hypergraphs. Edge-dependent node labels are unique to hypergraphs and cannot be easily expressed if we decompose hyperedges into pairwise edges between nodes.
Second, as shown in our experiments, using node and hyperedge embeddings obtained by existing hypergraph neural networks leads to limited performance
for the problem.
Lastly and most importantly, it has extensive real-world applications. For example, it can be used to predict the toxicity of chemical compounds (nodes) in specific reactions (hyperedges), offer guidance to students (nodes) who are expected to make limited contributions in group projects (hyperedges), and classify homonyms (nodes) based on the context of sentences (hyperedges). Moreover, the outputs of this problem (i.e., edge-dependent node properties) have proven to be useful in various applications.

In order to tackle the new classification task, we propose \method (\textbf{\underline{W}}ithin-\textbf{\underline{H}}yperedge \textbf{\underline{A}}ttention \textbf{\underline{T}}ran\textbf{\underline{s}}former \textbf{\underline{Net}}work). 
Most existing hypergraph neural networks do not explicitly consider edge-dependent relationships between node pairs within each hyperedge.
We design \WithinAttention, an attention mechanism where the edge-dependent embedding of each node is computed by attending to the other nodes in each hyperedge.
Inspired by the usefulness of positional encodings in the attention mechanism, we additionally use the centrality order of nodes within each hyperedge as the positional encoding, which makes attention in \WithinAttention even more edge-dependent.
The effectiveness of \method on the proposed task is demonstrated through extensive experiments.

Our contributions are summarized as follows:
\begin{itemize}[leftmargin=*]
    \item \textbf{New Problem}: To the best of our knowledge, we are the first to address the problem of classifying edge-dependent node labels.
    This problem is based on a unique property of hypergraphs and closely related to real-world applications.
    \item \textbf{Effective Model}: To tackle the problem, we design \method, a novel hypergraph neural network with attention and positional-encoding schemes that explicitly consider edge-dependent relationships between nodes.
    \item \textbf{Extensive Experiments}: Using six real-world hypergraphs, we demonstrate the superiority of \method over $10$ competitors on the considered task. We also show the usefulness of \method in three applications: ranking aggregation, node clustering, and  product-return prediction.
\end{itemize}
For reproducibility, we make the code and data available at \cite{github}.

The rest of this paper is organized as follows.
In Section~\ref{sec:related}, we review related works. In Section~\ref{sec:preliminaries}, we introduce preliminaries. In Section~\ref{sec:problem_formulation}, we describe our new problem with its applications. 
In Section~\ref{sec:proposed_approach}, we present \method, a hypergraph neural network for edge-dependent node classification. 
In Section~\ref{sec:evaluation}, we provide experimental results. Lastly, we make conclusions in Section~\ref{sec:conclusion}.

	\section{Related Work}
	\label{sec:related}
	In this section, we briefly survey related works on hypergraph neural networks and positional encodings.

\subsection{Hypergraph Neural Networks}

Hypergraph neural networks generalize graph neural networks into higher-order relationships in hypergraphs by allowing propagation between nodes through hyperedges.
Early models including HGNN~\cite{feng2019hypergraph} and HyperGCN~\cite{yadati2019hypergcn} first transform a given hypergraph into an ordinary graph  and then apply graph convolutions, losing high-order relationships embedded in hyperedges.
HNHN~\cite{dong2020hnhn} performs two steps of message passing using a nonlinear function to prevent equivalence to convolutions on the clique-expanded graph \footnote{A clique-expanded graph is an ordinary graph obtained by replacing each hyperedge with the clique composed by the constituent nodes.\label{footnote:clique}}. 
The first step updates hyperedge embeddings and the second updates node embeddings.
HNHN can vary weights on the contribution of incident embeddings during aggregation through hyperparameters.

HCHA~\cite{bai2021hypergraph} and HAT~\cite{hwang2021hyfer} apply attention mechanisms to improve the aggregation of the incident node or hyperedge embeddings.
The attention weights are calculated from the concatenation of node and hyperedge embeddings, which change during training.

UniGNN~\cite{huang2021unignn} and AllSet~\cite{chien2021you} generalize graph neural networks to hypergraphs.
UniGCNII is a special case of UniGNN that addresses over-smoothing, i.e.,  the convergence of embeddings of different nodes in deep (hyper)graph neural networks, by extending GCNII~\cite{chen2020simple} to hypergraphs.
AllSet is a framework composed of two multiset functions and generalizes most models, including HGNN, HyperGCN, HNHN, and HCHA.
AllSetTransformer replaces the multiset function with SetTransformer~\cite{lee2019set} and aggregates incident embeddings with attention to global learnable vectors.

A recent study with a similar motivation to ours presents a framework named HNN~\cite{aponte2022hypergraph} for jointly learning hyperedge embeddings and a set of hyperedge-dependent node embeddings. But
HNN makes the hyperedge-dependent node embeddings by simply concatenating the embeddings of nodes and incident hyperedges.

However, the importance of each node may vary depending on the other nodes it interacts with within each hyperedge, which may not be captured only by the relationship between the node and the hyperedge.
Most existing hypergraph neural networks do not explicitly consider edge-dependent relationships between pairs of nodes within each hyperedge.
Such relationships have been considered in 
a few studies in a way different from ours, 
but none of them directly apply to the considered problem.
HyperSAGNN \cite{zhang2019hyper} is designed for hyperedge prediction, and it uses two kinds of embeddings: (1) static embeddings, which are obtained from node features independently of hyperedges, and (2) dynamic embeddings,
which are calculated by aggregating the embeddings of the other nodes within hyperedges using pairwise attentions.
The model is trained to minimize the discrepancy between the static embedding and dynamic embedding of each node.
While HyperSAGNN uses pairwise attention, its goal is not to obtain hyperedge-dependent node representations.  
Higher-order Transformer~\cite{kim2021transformers} considers all subsets of nodes in each hyperedge,\footnote{$2^e$ for each hyperedge $e\subseteq \SV$, where $\SV$ is the set of nodes} and thus, computational and memory cost increases exponentially with the size of hyperedges.
The model has been applied only to uniform hypergraphs where the size of every hyperedge is identical, and it is non-trivial to apply it to real-world hypergraphs with hyperedges of varying sizes.

\subsection{Positional Encodings} \label{sec:pe}
It is known that many graph neural networks (GNNs) have difficulty in discriminating the positions of different nodes if the nodes share similar local structures~\cite{you2019position}. To improve the representation power, various positional encodings for graphs have been proposed. Such positional encodings, which we call \textit{absolute positional encodings}, are often added or concatenated to the node features, and they are based on random discriminating features~\cite{sato2021random,abboud2020surprising}, Weisfeiler–Lehman~\cite{zhang2020graph}, Laplacian Eigenvectors~\cite{dwivedi2020generalization,belkin2003laplacian,kreuzer2021rethinking}, random walk~\cite{li2020distance,dwivedi2021graph,ahmadi2020memory}, etc. 
Alternatively, GNNs that are able to capture positional information of nodes~\cite{you2019position,wang2022equivariant, dwivedi2021graph} can be used.

In Transformer-based GNNs, such positional encodings are usually used to calculate attention.
Graph Transformer~\cite{dwivedi2020generalization} and SAN~\cite{kreuzer2021rethinking} use Laplacian eigenvectors, while Graph-Bert~\cite{zhang2020graph} uses
the Weisfeiler algorithm~\cite{niepert2016learning}. Graphormer~\cite{ying2021transformers} adds the global degree centrality to node features to capture both semantic correlations and node centrality differences.
Its performance 
supports the effectiveness of employing node centrality as the positional encoding, which we also use in a different way for positional encoding.

Besides the absolute positional encodings, \textit{relative positional encodings}~\cite{raffel2020exploring, shaw2018self} are also used in Transformer-based GNNs to affect attention mechanisms based on the relative distance between a node pair (e.g., between a query and a key).
The relative distance can be computed based on positive definite kernels on graphs~\cite{mialon2021graphit}, learnable positional embeddings~\cite{ma2021graph}, shortest path distance~\cite{park2022grpe, ying2021transformers}, multi-step transition probability~\cite{zhao2021gophormer}, etc.
To the best of our knowledge, positional encodings specialized to hypergraphs have not been studied. 
In this work, we propose \PE, where the centrality order (i.e., ranking) of nodes within each hyperedge is used to encode the relative position within each hyperedge.

	\section{Preliminaries}
	\label{sec:preliminaries}
	In this section, we introduce concepts necessary to describe our method. The frequently-used symbols are summarized in Table~\ref{tab:notations}. 

\subsection{Hypergraphs}
A hypergraph $\Glong$ consists of a set of nodes $\SV =\{v_1, \ldots,v_N\}$ and a set of hyperedges $\SE = \{e_1, \ldots,e_M\} \subseteq2^{\SV}$. Each hyperedge $e\in \SE$ is a non-empty subset of $\SV$. 
We use $\mathcal{N}_{v} =\{e \in \SE :  v \in e\}$ to denote the set of hyperedges incident to the node $v$, i.e., the set of hyperedges that contain $v$.

\subsection{Attention Functions}
\smallsection{Scaled Dot-Product Attention}
For $n_q$ query vectors $\mathbf{Q} \in \mathbb{R}^{n_q \times d_k}$ and $n_k$ key-value vector pairs $\mathbf{K} \in \mathbb{R}^{n_k \times d_k}$ and $\mathbf{V} \in \mathbb{R}^{n_k \times d_v}$, where $d_k$ and $d_v$ represent vector sizes, the scaled dot-product attention computes the weighted sum of value vectors as follows:
\begin{equation*}
    Attention(\mathbf{Q}, \mathbf{K}, \mathbf{V}) = softmax(\frac{\mathbf{QK}^\top}{\sqrt{d_k}})\mathbf{V},
\end{equation*}
where $\sqrt{d_k}$ is typically used to avoid exploding gradients~\cite{vaswani2017attention}.
Note that the more similar a query vector and a key vector are (i.e., the larger their dot product is), the larger the corresponding weight is. 

\smallsection{Multihead Attention}
The multihead attention~\cite{vaswani2017attention} of dimension $d_k$ consists of $h$ attention modules of dimension $d_k/h$, whose outputs are concatenated for the final output, as follows:
\begin{equation*}
    MultiheadAttention(\mathbf{Q,K,V}) = Concat(\mathbf{O}_{1}, \cdots, \mathbf{O}_{h})\mathbf{W^{O}},
\end{equation*}
where $\mathbf{O}_i = Attention(\mathbf{Q}_i, \mathbf{K}_i, \mathbf{V}_i) = Attention(\mathbf{Q} \mathbf{W}^\mathbf{Q}_i, \mathbf{K} \mathbf{W}^\mathbf{K}_i, \mathbf{V} \mathbf{W}^\mathbf{V}_i)$.
That is,
$\mathbf{Q}$, $\mathbf{K}$, and $\mathbf{V}$ 
are mapped into subspaces 
$\mathbf{Q}_i\in \mathbb{R}^{n_q \times (d_k/h)}$, $\mathbf{K}_i\in \mathbb{R}^{n_k \times (d_k/h)}$, and $\mathbf{V}_i\in \mathbb{R}^{n_k \times (d_v/h)}$, respectively, by learnable parameters $\mathbf{W}^\mathbf{Q}_i\in \mathbb{R}^{d_k \times (d_k/h)}$ ,$\mathbf{W}^\mathbf{K}_i\in \mathbb{R}^{d_k \times (d_k/h)}$, and $\mathbf{W}^\mathbf{V}_i\in \mathbb{R}^{d_v \times (d_v/h)}$.
In our experiments,
the matrix $\mathbf{W}^\mathbf{O} \in \mathbb{R}^{d_v \times d_v}$ is fixed to the identity matrix, and the number of heads $h$ is fixed to $4$.

\begin{table}[t!]
	\caption{\label{tab:notations}Frequently-used symbols.}
	\scalebox{0.84}{
		\begin{tabular}{c|l}
			\toprule
			\textbf{Notation} & \textbf{Definition}\\
			\midrule
			$\Glong$ & a hypergraph with nodes $\SV$ and hyperedges $\SE$\\
			$N, M$ & the number of nodes and hyperedges \\
			$C$ & the number of (unique) edge-dependent node labels \\
			\midrule
			$v, e$ & a node $v \in \SV$ and a hyperedge $e \in \SE$ \\
			$y_{v, e}$ & the edge-dependent label of node $v$ in hyperedge $e$. \\
			$\mathcal{N}_{v}$ & the set of hyperedges incident to a node $v$ \\
			\midrule
			$l, m$ & the number of layers and inducing points\\
			$d_l$ & the hidden dimension at $l$-th layer \\
			$d_f$ & the dimension of \PE \\
			\midrule
			$\mathbf{F}$ & a node centrality feature matrix \\
			$\mathbf{I}_{w}$ & trainable inducing points in \WithinAttention \\
			$\mathbf{X}^{(l)}, \mathbf{H}^{(l)}$ & a node/hyperedge embedding matrix at layer $l$ \\
            $\mathbf{V}_{e}, \mathbf{E}_{v}$  & an embedding matrix of $e$ and $\mathcal{N}_{v}$ \\
			\bottomrule
		\end{tabular}}
\end{table}

\smallsection{Multihead Attention Block}\label{attblock}
We use the multihead attention block (MAB)~\cite{vaswani2017attention} as a component of our model.
It consists of a multihead attention module, a feed-forward layer, residual connections~\cite{he2016deep}, and layer normalization~\cite{ba2016layer}, as follows:
\begin{equation*}
    MAB(\mathbf{Q, K}) = LayerNorm(\mathbf{H} + FeedForward(\mathbf{H}))),
\end{equation*}
where $\mathbf{H} = LayerNorm(\mathbf{Q} + MultiheadAttention(\mathbf{Q,K,K}))$.
Note that the feed-forward layer does not change the dimensionality of the input matrix.

\smallsection{Attention with Inducing Points} \label{attwinducing} The scaled dot-product attention module requires all-pair dot-products between the query vectors and key vectors, making its time complexity quadratic, specifically $O(n_q n_k)$. 
SetTransformer~\cite{lee2019set} introduces an attention scheme that reduces the complexity to be linear.
With $m \ll \min(n_q, n_k)$ trainable inducing points, denoted by $I$, the all-pair dot-product $\MultiheadAtt(\mathbf{Q}, \mathbf{K})$ is approximated by $\MultiheadAtt(\mathbf{Q}, \MultiheadAtt(\mathbf{I, K}))$ whose time complexity is $O(m(n_q + n_k))$.
We adopt this strategy in our model for efficiency.

	\section{Problem Definition: Edge-dependent Node Classification in a Hypergraph}
	\label{sec:problem_formulation}
	We formally define the edge-dependent node classification problem and introduce its direct applications to real-world tasks.

\begin{figure*}[t]
    \vspace{-2mm}
    \begin{minipage}{0.51\linewidth}\label{fig:ouratt}
        \includegraphics[width=1.0\linewidth]{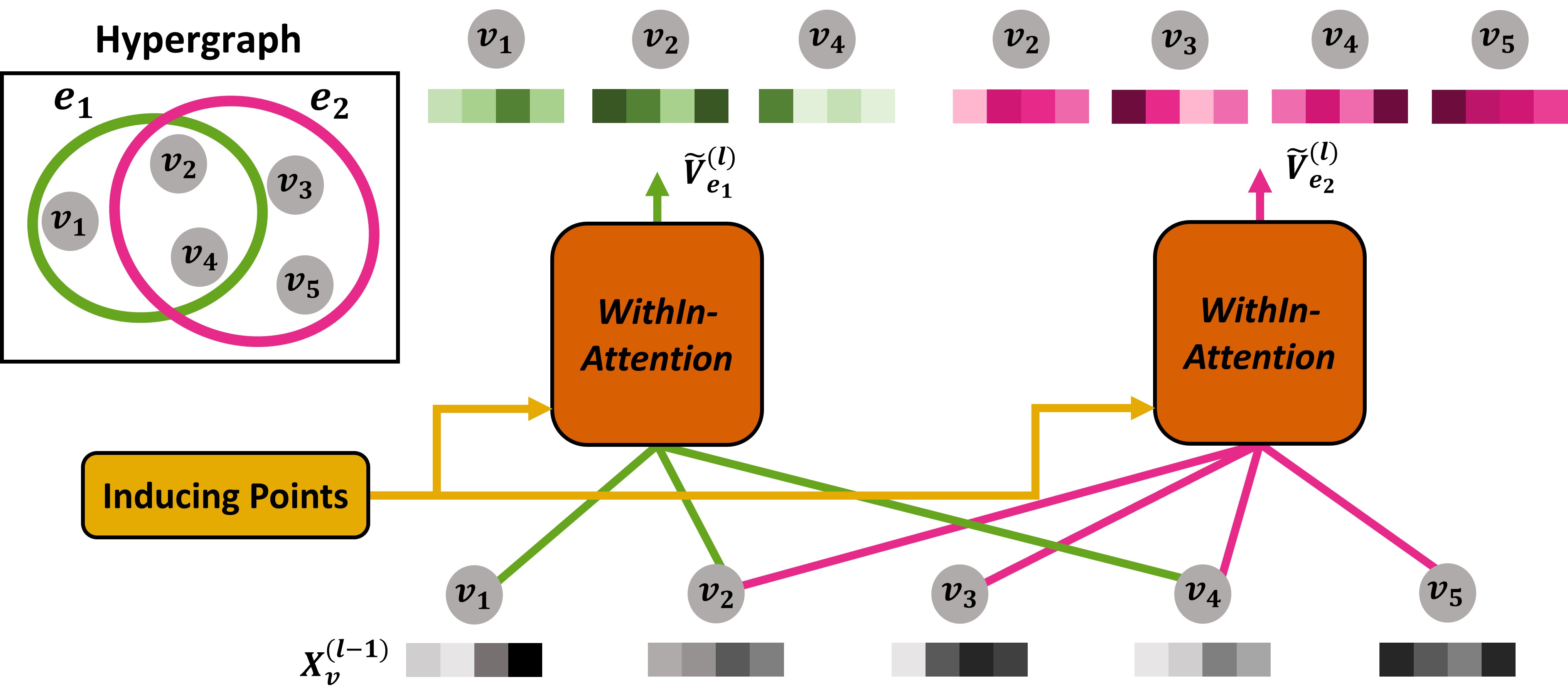}
        \subcaption{Illustration of \WithinAttention}
    \end{minipage}
    \begin{minipage}{0.48\linewidth}\label{fig:ourmodel}
         \includegraphics[width=1.0\linewidth]{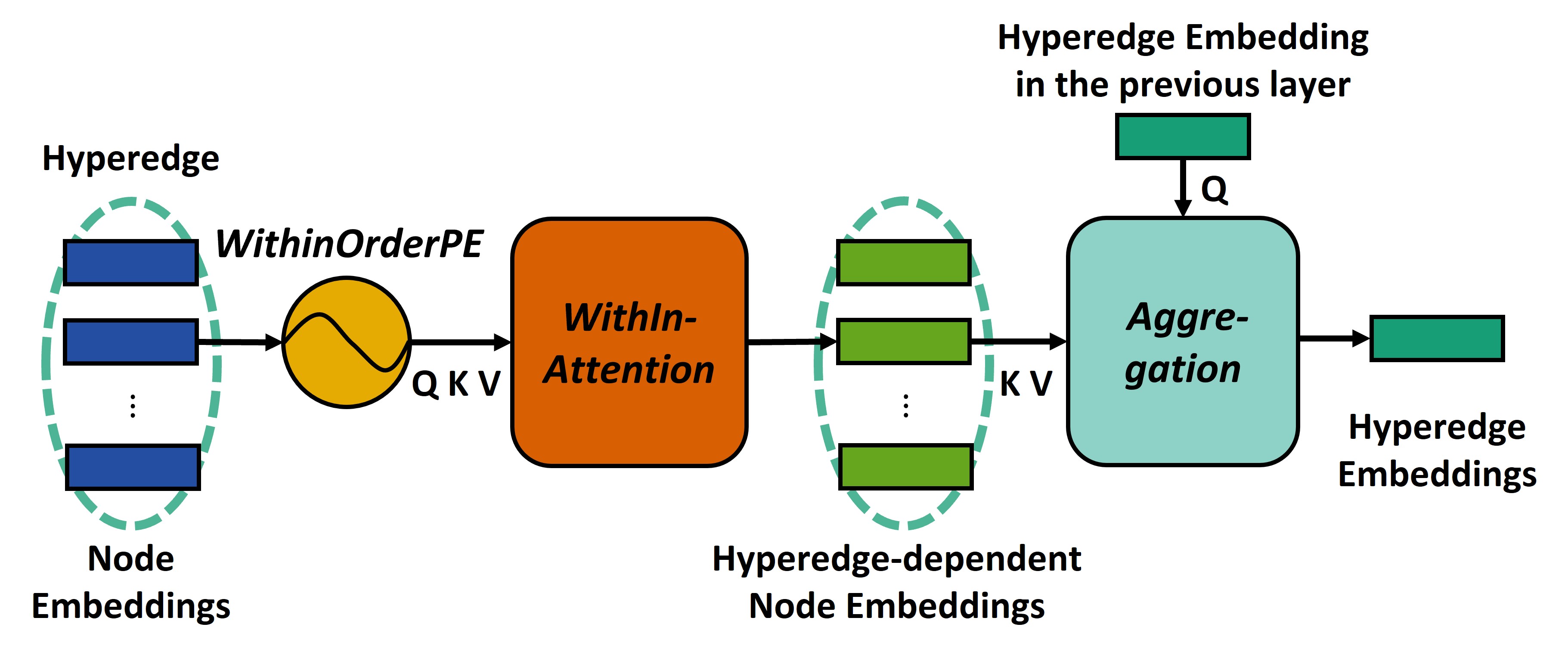}
         \vspace{0.1cm}
         \subcaption{Architecture of \method in Hyperedge Embedding Update} 
    \end{minipage}
    \caption{(a) In the given hypergraph, which consists of $e_1 = \{ v_1, v_2, v_4\}, e_2 = \{ v_2, v_3, v_4, v_5 \}$, \WithinAttention is applied to $e_1$ and $e_2$ independently. Even though the input feature of the node $v_2$ is the same, the output is different within $e_1$ and $e_2$. The inducing points for efficiently calculating pairwise relations are used globally. (b) The embeddings of the nodes in the target hyperedge are fed into \WithinAttention after \PE is added to each. Then, the outputs (i.e., edge-dependent node embeddings) are aggregated to update the embedding of the target hyperedge.}
\end{figure*}

\subsection{Problem Formulation}

In classical node classification, a single label is assigned to a node.
However, in hypergraphs, the labels of nodes can vary depending on the hyperedges that the nodes belong to.
For example, in a co-authorship hypergraph, the role of a researcher is categorized into the first, the last, and the middle author, and the position is likely to change in other  publications.
In this work, we introduce the edge-dependent node classification problem, formalized as follows.

\begin{pro}[Edge-dependent node classification]\label{problem}
Given (a) a hypergraph $\SG = (\SV, \SE)$, (b) a set of edge-dependent node labels 
in $\SE' \subset \SE$
(i.e., $y_{v,e}, \forall v \in e, \forall e \in \SE'$),
and optionally (c) a node feature matrix $\mathbf{X}$,
the problem is to correctly predict the unknown edge-dependent node labels in $\SE \setminus \SE'$ (i.e.,  $y_{v, e}, \forall v \in e, \forall e \in \SE \setminus \SE'$).
\end{pro}
\noindent This problem serves as an effective benchmark task and complements common hypergraph learning benchmarks for the following reasons: First, the edge-dependent node labels are unique to hypergraphs and cannot be reduced to node labels in the clique-expanded graphs (see Footnote~\ref{footnote:clique}). 
Thus, we can evaluate the capability of methods in capturing these unique characteristics of hypergraphs. 
Second, as demonstrated in the experiment section~\ref{sec:evaluation}, existing hypergraph neural networks have limitations in perfectly addressing this problem.
Lastly, the predictive outputs (i.e., edge-dependent node labels) can directly be applied to various applications.

\subsection{Applications}\label{benchmark_application}
Predicted edge-dependent node labels give information about the importance of nodes within each hyperedge, and leveraging this information has proven useful in a variety of applications.

\smallsection{Ranking Aggregation}
The task is to predict the overall (global) ranking of entities based on partial (local) rankings.
For example, in a multiplayer game where participants compete in matches, the aim of the task is to determine the global ranking of players using their local rankings from matches, enabling fair and competitive matches.
For this task, we can construct a hypergraph where (a) each node is a player, (b) each hyperedge is a match, (c) and the participants of a match are labeled based on their contribution levels during the match, as described in ~\cite{chitra2019random}. 
A similar task can be considered for a co-authorship dataset.
When the contribution level of each author in each publication is given, which is determined by the order of authors in each publication, the objective is to infer meaningful rankings for all authors.

\smallsection{Clustering}
Clustering refers to the problem of grouping similar elements, and the problem is commonly considered for data analysis.
Hayashi et al. \cite{hayashi2020hypergraph} designed a node clustering algorithm that leverages edge-dependent node weights, demonstrating its high-quality clusters for real-world hypergraphs.
Edge-dependent node labels, inferred by resolving  Problem~\ref{problem}, can be directly transformed into node weights for clustering purposes.
For example, in a co-authorship dataset, the weights of authors (nodes), determined by their positions (e.g., first, last, or others) within each publication (hyperedge), can be utilized to enhance clustering results.

\smallsection{Product Return Prediction}
In E-commerce, predicting customers' intention to return products is helpful to take proactive actions. 
Li et al.~\cite{li2018tail} showed that the product return probability of each target basket can be accurately estimated by using hypergraphs where nodes and hyperedges are baskets and products, respectively, and baskets are labeled based on the number of each product they contain. 
The authors showed that these edge-dependent node labels are informative in predicting product returns. For instance, multiple purchases of the same product within a basket may indicate that the customer purchases the same product with different colors or sizes, increasing the likelihood of product return.

	\section{Proposed Model: \method}
	\label{sec:proposed_approach}
	


We propose \method (\textbf{W}ithin-\textbf{H}yperedge \textbf{A}ttention \textbf{T}ran\textbf{s}former \textbf{Net}work) for edge-dependent node classification,
 with each layer consisting of two message passing steps:
(1) updating hyperedge embeddings by aggregating node embeddings belonging to each hyperedge, and (2) updating node embeddings by aggregating hyperedge embeddings that include the corresponding node.
In each step, input embeddings of nodes (or hyperedges) are adapted using an attention module called \WithinAttention, and then aggregated into hyperedge (or node) embeddings.
\WithinAttention considers edge-dependent (or node-dependent) relations between adjacent nodes (or hyperedges) by incorporating edge-dependent positional encodings named \PE. 
The embeddings generated from the final layer are utilized to predict edge-dependent node labels.

\subsection{\WithinAttention: Attention to Other Nodes within Hyperedges}\label{method:attention:within} 

In many real-world hypergraphs, the importance or role of a node is shaped by the other nodes within the same hyperedge. 
For instance, if a grown-up researcher co-authors a paper only with students, she is likely to be the last author of the paper. 
Conversely, if she co-authors with other grown-up researchers, then the probability of her being the last author of that paper is relatively low. 

Motivated by this observation, we devise \WithinAttention based on  Transformer~\cite{vaswani2017attention}, and it adapts a node embedding by attending to the other nodes in the same hyperedge.
Specifically, it uses a set of node embeddings as queries, keys, and values in the attention mechanism. 
It, thereby, models and utilizes relations between nodes by taking the dot-product of all node pairs within the hyperedge.

However, calculating the dot-product for every node pair has  computational complexity quadratic in the hyperedge size,
 making it challenging to use in large-scale real-world hypergraphs.
To overcome this issue, we adopt the inducing point method (described in Section~\ref{attwinducing}) in SetTransformer~\cite{lee2019set}, 
which performs comparably to all-pair attention while being significantly more efficient.

Let $\mathbf{V}_{e}^{(l)} = \{ \mathbf{X}_u^{(l)} : u \in e \}$ denote the set of the embeddings of nodes in a hyperedge $e$ in the $l$-th layer, which are in the form of a matrix in $\mathbb{R}^{ |e| \times d_l}$. 
Along with $\mathbf{V}_{e}^{(l)}$, \WithinAttention uses $m$ number of $d_l$-dimensional trainable inducing points $\mathbf{I}_w \in \mathbb{R}^{m \times d_l}$, where $m$ is typically much smaller than $\max_{e \in \SE} |e|$. In our experiments, $m$ is fixed to $4$.
Then, \WithinAttention is formally expressed as follows:
\begin{equation}
    \WithinAttention(\mathbf{V}_{e}^{(l)} ; \mathbf{I}_w ) = \MultiheadAtt(\mathbf{V}_{e}^{(l)}, \MultiheadAtt(\mathbf{I}_w, \mathbf{V}_{e}^{(l)})).
\end{equation}
Even though $\mathbf{I}_w$ is shared across
all hyperedges, training the inducing points can be seen as finding a proper projection function for any input onto a lower-dimensional space.
Additionally, the inner \MultiheadAtt (defined in Section~\ref{attwinducing}) would summarize the node embeddings in $e$ using the well-trained $\mathbf{I}_w$. Through these outputs, the outer \MultiheadAtt would generate edge-dependent representations of nodes within $e$, leveraging the attention 
between the input node embeddings and the summary of them. 
Consequently, the resulting representations serve
as an approximation of the attention between all node pairs while avoiding the quadratic complexity.

\subsection{\PE: Using Centrality for Positional Encoding}\label{method:attention:rank}

We expect that edge-dependent node labels are closely related to the relative positions of nodes within each hyperedge, which can be measured, for example, by centrality relative to the other nodes in each hyperedge. 
For instance, in co-authorship datasets, degree centrality indicates the number of papers authored by each researcher.
Hence, the author with the highest centrality among all authors of a paper is more likely to be the last author.

Based on this reasoning, we devise \PE, a positional encoding that facilitates edge-dependent attention between nodes within each hyperedge.
Specifically, we use the relative order of node centrality within the hyperedge for \PE.
We define the \textit{order} of each element $a$ in a set $A$ as follows:
\begin{equation}\label{equation:order}
    Order(a, A) = \sum\nolimits_{a' \in A} \mathds{1}(a' \leq a).
\end{equation}
Then, given node centralities $\mathbf{F} \in \mathbb{R}^{N \times d_f}$ where $d_f$ is the number of centrality measures, corresponding to the dimensionality of positional encodings,
we define \PE of a node $v$ within a hyperedge $e$ as follows: 
\begin{equation}\label{equation:rank}
    \PE(v, e ; \mathbf{F}) = \parallel_{i=1}^{d_f} \frac{1}{|e|} Order(\mathbf{F}_{v,i}, \{ \mathbf{F}_{u,i} : \forall u \in e \}),
\end{equation}
where $\parallel$ represents the concatenation of the orders with respect to different centrality measures. 
We simply add \PE to the node embeddings and feed them into \WithinAttention. In our experiments, we leverage four types of node centrality: 
degree, eigenvector centrality, PageRank, and coreness. Details of them are provided in Appendix ~\ref{appendix:abl_cent}.

Similarly, for the message passing from hyperedges to nodes, we explicitly give the target node's order within the source hyperedge.
In other words, the same position encoding is used for each hyperedge-node pair in both directions of message passing.
This is because the role of a hyperedge $e$ with respect to the target node $v \in e$ is also affected by the importance of node $v$ in $e$.


\begin{table*}[t]
    \vspace{-2mm}
	\begin{center}
		\caption{\label{tab:datasets} Summary of real-world hypergraphs with edge-dependent labels.}
		\scalebox{0.84}{
		    \begin{tabular}{l| r  r  r  r  r | r r r c c}
				\toprule
				Dataset & \bigcell{c}{Num. of\\Nodes} & \bigcell{c}{Num. of\\Hyperedges} & \bigcell{c}{Max. of\\Node Deg.} & \bigcell{c}{Max. of\\Hyperedge Size} & \bigcell{c}{Sum of\\Hyperedge Size} & \bigcell{c}{Num. of\\Class 0} & \bigcell{c}{Num. of\\Class 1} & \bigcell{c}{Num. of\\Class 2} & \bigcell{c}{Corr. \\w/ Centrality} & \bigcell{c}{Avg.\\Entropy}\\ 
				\midrule
				Coauth-DBLP & 108,484 & 91,266 & 236 & 36 & 321,011 & 91,266 & 138,479 & 91,266 & 0.19 & 0.13\\ 
				Coauth-AMiner & 1,712,433 & 2,037,605 & 752 & 115 & 5,129,998 & 2,037,605 & 1,652,332 & 1,503,061 & 0.24 & 0.13\\
				Email-Enron & 21,251 & 101,124 & 18,168 & 948 & 1,186,521 & 635,268 & 450,129 & 101,124 & 0.10 & 0.28\\
				Email-Eu & 986 & 209,508 & 8,659 & 59 & 541,842 & 209,508 & 332,334 & - & 0.24 & 0.48 \\
				Stack-Biology & 15,490 & 26,823 & 1,318 & 12 & 56,257 & 26,290 & 18,444 & 11,523 & 0.29 & 0.10\\
				Stack-Physics & 80,936 & 200,811 & 6,332 & 48 & 479,809 & 194,575 & 201,121 & 84,113 & 0.30 & 0.12\\
				\bottomrule
			\end{tabular}}
	\end{center}
\end{table*}

\subsection{\method: Our Final Model}\label{method:overall} 

We introduce \method, a hypergraph neural network that integrates the afore-described modules.
In each layer, hyperedge embeddings are first updated and then node embeddings are updated. Before updating hyperedge embeddings, edge-dependent node embeddings are computed by incorporating edge-dependent relationships between nodes within each hyperedge via \WithinAttention (Eq.~\eqref{equation:method_att}) and \PE (Eq.~\eqref{equation:method_pe}). 
The hyperedge embedding is then updated by aggregating these edge-dependent node embeddings with weights determined by the dot-product with the hyperedge embedding from the previous layer. This process can be simplified using \MultiheadAtt (defined in Section ~\ref{attblock}) (Eq.~\eqref{equation:method_agg}).
Specifically, a hyperedge $e$'s embedding $\mathbf{H}_e^{(l)}$ is updated as follows:
\begin{equation}\label{equation:method_pe}
    \mathbf{V}_{e,\boxplus}^{(l)} = \{ \mathbf{X}_v^{(l-1)} \boxplus \PE(v, e)) : v \in e  \},
\end{equation}
\begin{equation}\label{equation:method_att}
    \tilde{\mathbf{V}}_{e}^{(l)} = \WithinAttention(\mathbf{V}_{e,\boxplus}^{(l)}),
\end{equation}
\begin{equation}\label{equation:method_agg}
    \mathbf{H}_e^{(l)} = \MultiheadAtt(\mathbf{H}_e^{(l-1)},\tilde{\mathbf{V}}_{e}^{(l)})
\end{equation}
The notation $\boxplus$ denotes the addition of positional encodings and embeddings, after matching the dimension of the former with that of the latter through a learnable weight matrix, $\mathbf{W} \in \mathbb{R}^{d_{f} \times d_{l-1}}$~\footnote{Formally, $\mathbf{A} \boxplus \mathbf{B} = \mathbf{A} + \mathbf{B} \mathbf{W}$ where $\mathbf{A} \in \mathbb{R}^{n \times d_{A}}$, $\mathbf{B} \in \mathbb{R}^{n \times d_{B}}$ and $\mathbf{W} \in \mathbb{R}^{d_{B} \times d_{A}}$.}.

Similarly, node embeddings are updated by aggregating (Eq.~\eqref{equation:method_agg_e}) node-dependent hyperedge embeddings from \WithinAttention (Eq.~\eqref{equation:method_att_e}) and \PE (Eq.~\eqref{equation:method_pe_e}).
Specifically, a node $v$'s embedding $\mathbf{X}_v^{(l)}$ is updated as follows:
\begin{equation}\label{equation:method_pe_e}
    \mathbf{E}_{v,\boxplus}^{(l)} = \{ \mathbf{H}_e^{(l)} \boxplus \PE(v, e)) : e \in \mathcal{N}_{v}  \},
\end{equation}
\begin{equation}\label{equation:method_att_e}
    \tilde{\mathbf{E}}_{v}^{(l)} = \WithinAttention(\mathbf{E}_{v,\boxplus}^{(l)}),
\end{equation}
\begin{equation}\label{equation:method_agg_e}
    \mathbf{X}_v^{(l)} = \MultiheadAtt(\mathbf{X}_v^{(l-1)},\tilde{\mathbf{E}}_{v}^{(l)})
\end{equation}

One can stack more than one \WithinAttention module before aggregation to make attention between nodes within each hyperedge even more edge-dependent.
We stack two \WithinAttention modules in our experiments.

\subsection{Application to Edge-Dependent Node Classification}\label{method:classifier}

Given node and hyperedge embeddings, we predict edge-dependent node labels with a single-layer perceptron classifier, denoted as $\psi$, which takes the concatenation of node and hyperedge embeddings as its input.
Let $\mathbf{X}^{(L)}$ and $\mathbf{H}^{(L)}$ be the node and hyperedge embeddings generated from the last $L$-th layer of \method respectively.
Then, the label of a node $v$ in a hyperedge $e$ is predicted as follows:
\begin{equation}\label{eq:classifier}
    \hat{y}_{(v, e)} = \argmax( \psi([\mathbf{X}_v^{(L)} \cat \mathbf{H}_e^{(L)}]) ).
\end{equation}
Eq.~\eqref{eq:classifier} 
can readily be applied to other hypergraph neural networks, such as HNHN~\cite{dong2020hnhn}, that generate separate node and hyperedge embeddings as their outputs.
Thus, any accuracy gain of \method with Eq.~\eqref{eq:classifier}, compared to such models, is attributed to its ability to generate more informative node and hyperedge embeddings.

Alternatively, one can utilize intermediate edge-dependent node embeddings 
\smash{$\tilde{\mathbf{V}}_{e}^{(L)} \in \mathbb{R}^{|e| \times {d_L}}$}
 in Eq.~\eqref{equation:method_att}, which are generated inside a \method layer:
\begin{equation}\label{eq:classifier2}
    \bar{y}_{(v, e)} = \argmax ( \psi( \tilde{\mathbf{V}}_{e}^{(L)}[v] ))
\end{equation}
where $[v]$ means indexing the vector corresponding to node $v$.
In this way, one can avoid conducting the explicit concatenation with hyperedge embeddings, since \smash{$\tilde{\mathbf{V}}_{e}^{(L)}$} already has such information.

Table \ref{tab:classifier} in Appendix \ref{appendix:additional_abl} compares the two approaches of Eq.~\eqref{eq:classifier} and \eqref{eq:classifier2}.\footnote{We remove \PE from the last layer of \method when applying the approach of Eq.~\eqref{eq:classifier2}, because if not, the node centralities affect the classification performance too much.}
There is no clear superiority between the two choices, and both of them consistently outperform the best competitor.
Still, we choose the first approach as our default choice because it is more general and memory-efficient: we can store the node and hyperedge embeddings separately and then concatenate them for any downstream task utilizing edge-dependent node embeddings.\footnote{which is $O(N+M)$ instead of $O(NM)$ when $N$ nodes and $M$ hyperedges exist.}

\subsection{Complexity Analysis}\label{method:analysis}
We analyze the time and space complexities of \method.

\smallsection{Time Complexity}
\WithinAttention requires dot products between the input embeddings and inducing points, and the aggregation involves dot products between the input embeddings and a single query. 
Thus, the total time complexity of the inference step of \method without \PE is $O(\sum_{l=1}^{L}(\sum_{e \in \SE} (|e|md_{l}) + \sum_{v \in \SV} (|N_v|md_{l}))) = O(m\sum_{l=1}^{L}d_{l} \sum_{e \in \SE} |e|)$, where the last equality is from $\sum_{v \in \SV} |\mathcal{N}_{v}|$ $=\sum_{e\in \SE} |e|$.
As for \PE, we first compute four node centralities of each node (spec., degree, coreness, eigenvector centrality, and PageRank), whose complexity is $O(\sum_{e\in \SE} |e|)$ (see Appendix~\ref{appendix:complexity} for details).
Next, we compute the order of nodes within hyperedges by sorting nodes based on each centrality value, whose time complexity is $O(\sum_{e \in \SE} |e|\log|e|)$.
Thus, the complexity of computing \PE is $O(\sum_{e \in \SE} |e| \log |e|)$, and the total time complexity of \method with \PE is $O(m\sum_{l=1}^{L}d_{l} \sum_{e \in \SE} |e|+ \sum_{e\in \SE} |e|\log|e|)$.

\smallsection{Space Complexity}
We consider full-batch training.
Storing the input hypergraph $\SG = (\SV, \SE)$ and the $d_f$-dimensional positional encodings obtained by \PE requires $O(d_f \sum_{e \in \SE} |e|)$ space.
We first consider the message passing from nodes to hyperedges at the $l$-th layer.
\WithinAttention requires (a) node embeddings, whose size $O(|V|d_{l})$, (b) a weight matrix for adjusting the dimension of positional encodings, whose size is $O(d_{f}d_{l})$, and (c) $m$ inducing points of $d_{l}$ dimension, whose size is $O(md_{l})$. It also requires two attention matrices between the inputs and $m$ inducing points, whose sizes are $|e|$ by $m$ and $m$ by $|e|$ for each hyperedge $e$.
Thus, the additional space required for the node-to-hyperedge message passing is $O(md_{l} + d_{f}d_{l} + |V|d_{l} + m\sum_{e\in \SE} |e|)$.

Similarly, the message passing from hyperedges to nodes at the $l$-th layer requires $O(md_{l} + d_{f}d_{l}+ |E|d_{l} + m\sum_{v \in \SV} |\mathcal{N}_{v}|)$.
Since $\sum_{v \in \SV} |\mathcal{N}_{v}|=\sum_{e\in \SE} |e|$, the total space complexity of \method 
with $L$ layers is $O( (m + d_{f} + |\SV| + |\SE|) \sum_{l=1}^{L} d_{l} + (Lm + d_{f}) \sum_{e\in \SE}|e| ) $.




\subsection{Comparison to Existing Models}\label{method:comp}

To efficiently capture the edge-dependent relationships between node pairs within hyperedges, we incorporate the encoder part of SetTransformer~\cite{lee2019set} as a component of the message passing in our model. It is important to note that SetTransformer is primarily designed to obtain proper representations of sets, and its primary focus is not on hypergraph representation learning.

While both \method and SetTransformer aggregate input embeddings using a weighted sum, there are notable differences in the aggregation methods. SetTransformer uses weights obtained by attending to a global learnable parameter, whereas \method attends to the previous embedding of the target. Additionally, \method incorporates \PE into attention, enhancing its ability to capture edge-dependent relationships within hyperedges.

There is another approach, called AllSetTransformer~\cite{chien2021you}, which integrates SetTransformer into hypergraph neural networks. However, it should be noted that AllSetTransformer does not adopt the encoder part of SetTransformer but instead adopts its decoder with modifications to the number of stacking feed-forward layers in the multi-head attention block. Therefore, the usage of SetTransformer in AllSetTransformer differs from how \method utilizes it.

	\section{Experimental Results}
	\label{sec:evaluation}
\begin{table*}[t]
    \vspace{-2mm}
    \begin{center}
        \caption{\label{tab:expresult} \underline{\smash{Results of Edge-Dependent Node Classification}}: Mean and standard deviation of Micro-F1 and Macro-F1 scores over five independent runs are reported. Best scores are in \textbf{bold}. Note that \method performs best in all datasets.}
        \scalebox{0.7}{
            \begin{tabular}{l l | c c | c c c c c c | c c || c}
            \toprule
            Dataset & Metric & BaselineU & BaselineP & HNHN & HGNN & HCHA & HAT & UniGCNII & HNN & HST & AST & \method\\ 
            \midrule
            \multirow{2}{*}{\shortstack[l]{Coauth-\\DBLP}} & MicroF1 & 0.333 $\pm$ 0.001 & 0.346 $\pm$ 0.001 & 0.486 $\pm$ 0.004  & 0.540 $\pm$ 0.004  & 0.451 $\pm$ 0.007  & 0.503 $\pm$ 0.004  & 0.497 $\pm$ 0.003  & 0.488 $\pm$ 0.006  & 0.564 $\pm$ 0.004  & 0.495 $\pm$ 0.038  & \textbf{0.605} $\pm$ 0.002  \\ 
            & MacroF1 & 0.330 $\pm$ 0.001 & 0.332 $\pm$ 0.001 & 0.478 $\pm$ 0.008  & 0.519 $\pm$ 0.002  & 0.334 $\pm$ 0.048  & 0.483 $\pm$ 0.006  & 0.476 $\pm$ 0.002  & 0.482 $\pm$ 0.006  & 0.549 $\pm$ 0.003  & 0.487 $\pm$ 0.040  & \textbf{0.595} $\pm$ 0.002  \\ 
            \midrule
            \multirow{2}{*}{\shortstack[l]{Coauth-\\AMiner}} & MicroF1 & 0.334 $\pm$ 0.000 & 0.339 $\pm$ 0.000 & 0.520 $\pm$ 0.002  & 0.566 $\pm$ 0.002  & 0.468 $\pm$ 0.020  & 0.543 $\pm$ 0.002  & 0.520 $\pm$ 0.001  & 0.543 $\pm$ 0.002 & 0.596 $\pm$ 0.007  & 0.577 $\pm$ 0.005  & \textbf{0.630} $\pm$ 0.005  \\ 
            & MacroF1 & 0.332 $\pm$ 0.000 & 0.333 $\pm$ 0.000 & 0.514 $\pm$ 0.002  & 0.551 $\pm$ 0.004  & 0.447 $\pm$ 0.040  & 0.533 $\pm$ 0.003  & 0.507 $\pm$ 0.001  & 0.533 $\pm$ 0.002  & 0.583 $\pm$ 0.008  & 0.570 $\pm$ 0.002  & \textbf{0.623} $\pm$ 0.007  \\ 
            \midrule
            \multirow{2}{*}{\shortstack[l]{Email-\\Enron}} & MicroF1 & 0.334 $\pm$ 0.001 & 0.439 $\pm$ 0.001 & 0.738 $\pm$ 0.028  & 0.725 $\pm$ 0.004  & 0.666 $\pm$ 0.010  & 0.817 $\pm$ 0.001  & 0.734 $\pm$ 0.010  & 0.763 $\pm$ 0.003  & 0.779 $\pm$ 0.067  & 0.796 $\pm$ 0.014  & \textbf{0.826} $\pm$ 0.001  \\ 
            & MacroF1 & 0.300 $\pm$ 0.001 & 0.333 $\pm$ 0.001 & 0.637 $\pm$ 0.023  & 0.674 $\pm$ 0.003  & 0.464 $\pm$ 0.002  & 0.753 $\pm$ 0.004  & 0.656 $\pm$ 0.010  & 0.679 $\pm$ 0.007  & 0.681 $\pm$ 0.123  & 0.719 $\pm$ 0.020  & \textbf{0.760} $\pm$ 0.004  \\ 
            \midrule
            \multirow{2}{*}{\shortstack[l]{Email-\\Eu}} & MicroF1 & 0.500 $\pm$ 0.001 & 0.525 $\pm$ 0.001 & 0.643 $\pm$ 0.004  & 0.633 $\pm$ 0.001  & 0.620 $\pm$ 0.000  & 0.669 $\pm$ 0.001  & 0.630 $\pm$ 0.005  & OutOfMemory  & \textbf{0.671} $\pm$ 0.001  & 0.666 $\pm$ 0.005  & \textbf{0.671} $\pm$ 0.000  \\ 
            & MacroF1 & 0.493 $\pm$ 0.001 & 0.499 $\pm$ 0.001 & 0.552 $\pm$ 0.014  & 0.533 $\pm$ 0.008  & 0.497 $\pm$ 0.001  & 0.638 $\pm$ 0.002  & 0.565 $\pm$ 0.013  & OutOfMemory  & 0.640 $\pm$ 0.002  & 0.624 $\pm$ 0.021  & \textbf{0.646} $\pm$ 0.003  \\ 
            \midrule
            \multirow{2}{*}{\shortstack[l]{Stack- \\Biology}} & MicroF1 & 0.335 $\pm$ 0.000 & 0.368 $\pm$ 0.001 & 0.640 $\pm$ 0.005  & 0.689 $\pm$ 0.002  & 0.589 $\pm$ 0.007  & 0.661 $\pm$ 0.005  & 0.610 $\pm$ 0.004  & 0.618 $\pm$ 0.015  & 0.694 $\pm$ 0.002  & 0.571 $\pm$ 0.054  & \textbf{0.742} $\pm$ 0.003  \\  
            & MacroF1 & 0.326 $\pm$ 0.000 & 0.334 $\pm$ 0.003 & 0.592 $\pm$ 0.006  & 0.624 $\pm$ 0.007  & 0.465 $\pm$ 0.060  & 0.606 $\pm$ 0.005  & 0.433 $\pm$ 0.007  & 0.568 $\pm$ 0.013  & 0.631 $\pm$ 0.006  & 0.446 $\pm$ 0.081  & \textbf{0.686} $\pm$ 0.004  \\ 
            \midrule
            \multirow{2}{*}{\shortstack[l]{Stack- \\Physics  }} & MicroF1 & 0.333 $\pm$ 0.001 & 0.370 $\pm$ 0.000 & 0.506 $\pm$ 0.053  & 0.686 $\pm$ 0.004  & 0.622 $\pm$ 0.003  & 0.708 $\pm$ 0.005  & 0.671 $\pm$ 0.022  & 0.683 $\pm$ 0.005  & 0.755 $\pm$ 0.010  & 0.728 $\pm$ 0.039  & \textbf{0.770} $\pm$ 0.003  \\ 
            & MacroF1 & 0.322 $\pm$ 0.001 & 0.332 $\pm$ 0.000 &  0.422 $\pm$ 0.043  & 0.630 $\pm$ 0.002  & 0.481 $\pm$ 0.007  & 0.643 $\pm$ 0.009  & 0.492 $\pm$ 0.016  & 0.617 $\pm$ 0.005  & 0.666 $\pm$ 0.013  & 0.646 $\pm$ 0.046  & \textbf{0.707} $\pm$ 0.004  \\ 
            \bottomrule
        \end{tabular}}
    \end{center}
\end{table*}

We evaluate \method by answering the following questions:

\begin{itemize}[leftmargin=*]
    \item \textbf{Q1.} Does \method accurately predict the edge-dependent labels of nodes?
    \item \textbf{Q2.} Does \method classify the same node  differently depending on hyperedges? 
    \item \textbf{Q3.} Does each component of \method make a meaningful contribution to the performance?
    \item \textbf{Q4.} How can we apply \method to real-world downstream tasks? Does \method show usefulness in these tasks?
\end{itemize}

\subsection{Experimental settings}
\label{sec:exp:setting}

\smallsection{Datasets}\label{datasets}
We use six real-world hypergraphs from three domains:
\begin{itemize}[leftmargin=*]
    \item \textbf{Co-authorship} (DBLP~\cite{Swati2016dblp} and AMinerAuthor\footnote{https://www.aminer.org/aminernetwork}): Each hyperedge indicates a publication, and the nodes in it are the authors of the publication. Edge-dependent node labels indicate the orders of authors (first, last, or others) in each publication. 
    We exclude publications where the authors are in alphabetical order, which may not reflect the difference in the authors' contributions.
    \item \textbf{Email} (Enron\footnote{https://www.cs.cmu.edu/~enron/} and Eu~\cite{paranjape2017motifs}): Each hyperedge represents an email, and the nodes in it represent the people involved in the email. Edge-dependent labels distinguish (1) the sender, (2) the receivers, and (3) the CC'ed.
    \item \textbf{StackOverflow} (Biology\footnote{\label{footnote:stack}https://archive.org/download/stackexchange} and Physics\footref{footnote:stack}): Each hyperedge represents a post, and the nodes in it are the users contributing to this post. Edge-dependent labels distinguish (1) the questioner, (2) answerers chosen by the  questioner, and (3) the other answerers.
\end{itemize}
Table~\ref{tab:datasets} shows some statistics for the datasets. We measure the correlation coefficient between within-edge node-centrality orders and edge-dependent node labels, using the average of Cramer's coefficient matrix~\cite{cramer1946mathematical}. 
We also report the average entropy in each node's label distribution. Note that the average entropy is non-zero in all datasets, implying that node labels do vary across hyperedges. 

\smallsection{Competitors}
We implemented all models using Deep Graph Library (DGL) ~\cite{wang2019deep}.
In all competitors, edge-dependent labels are predicted by a single-layer perceptron from the concatenation of the global embeddings of nodes and hyperedges, as in ours (Eq.~\eqref{eq:classifier}).
As competitors, we use seven hypergraph neural networks that generate both hyperedge and node embeddings and thus can be used for edge-dependent node classification: HNHN~\cite{dong2020hnhn}, HGNN~\cite{feng2019hypergraph}, HCHA~\cite{bai2021hypergraph}, HNN~\cite{aponte2022hypergraph}, HAT~\cite{hwang2021hyfer}, UniGCNII~\cite{huang2021unignn}, and AllSetTransformer (AST) ~\cite{chien2021you}.
We do not include HyperGCN~\cite{yadati2019hypergcn} as a competitor because it does not provide hyperedge embeddings. 
In addition, we consider HypergraphSetTransformer (HST), which extends SetTransformer~\cite{lee2019set} to hypergraphs by applying it to obtain the representations of sets of incident nodes and hyperedges in the two stages of message passing.
Lastly, we consider two simple approaches for comparison: (a) \textit{BaselineU}, which predicts node labels uniformly at random, and (b) \textit{BaselineP}, which randomly assigns labels proportionally to their global distribution. 

\smallsection{Other Settings}
Since external features are absent, we create initial node features by adopting 2nd-order random walks on hypergraphs as in \cite{zhang2019hyper}. 
Specifically, we construct fixed-length random walks for each node and obtain embeddings using a skip-gram model~\cite{mikolov2013efficient,mikolov2013distributed}. 
All parameters are initialized via Xavier initialization~\cite{he2015delving} and trained with Adam optimizer~\cite{kingma2014adam}.
\PE incorporates four centrality measures: degree, coreness, eigenvector centrality, and PageRank (refer to Appendix~\ref{appendix:abl_cent}).
We randomly divide all hyperedges in each hypergraph into training (60\%), validation (20\%), and test (20\%) sets.
For the edge-dependent node classification task, we stop training when the number of epochs reaches $100$ or the accuracy on the validation set no longer improves for $25$ epochs. 
Whereas, for downstream tasks, we stop training when the number of epochs reaches $300$ or the mean of Micro-F1 and Macro-F1 scores does not change for $10$ epochs.
We use the search space described in Appendix~\ref{appendix:hp} for hyperparameter tuning. 
We select the hyperparameter values that maximize the mean of Micro-F1 and Macro-F1 scores on the validation set and report its performance on the test set over five independent runs. 

\begin{table*}[t]
    \vspace{-2mm}
    \begin{center}
        \caption{\label{tab:jsdresult} \underline{\smash{Jensen-Shannon divergence (JSD) between Ground-truth and Predicted Node Label Distributions}}: The average and standard deviation of the average JSD over five independent runs are reported.
        The lower JSD is, the better a model preserves node-level label distributions.
        Best scores are in \textbf{bold}.
        \method yields the lowest JSD in all datasets.}
        \scalebox{0.71}{
            \begin{tabular}{l| c c | c c c c c c | c c || c}
            \toprule
            Dataset & BaselineU & BaselineP & HNHN & HGNN & HCHA & HAT & UniGCNII & HNN & HST & AST & \method\\
            \midrule
            Coauth-DBLP & 0.532 $\pm$ 0.002 &  0.518 $\pm$ 0.002 &  0.450 $\pm$ 0.002 &  0.394 $\pm$ 0.005 &  0.450 $\pm$ 0.002 &  0.429 $\pm$ 0.004 &  0.449 $\pm$ 0.006 &  0.450 $\pm$ 0.003 &  0.388 $\pm$ 0.006 &  0.453 $\pm$ 0.038 &  \textbf{0.350} $\pm$ 0.002 \\
            Coauth-AMiner & 0.529 $\pm$ 0.000 &  0.523 $\pm$ 0.000 &  0.440 $\pm$ 0.003 &  0.372 $\pm$ 0.002 &  0.462 $\pm$ 0.002 &  0.414 $\pm$ 0.003 &  0.424 $\pm$ 0.002 &  0.411 $\pm$ 0.003 &  0.356 $\pm$ 0.005 &  0.382 $\pm$ 0.009 &  \textbf{0.328} $\pm$ 0.004 \\ 
            Email-Enron & 0.486 $\pm$ 0.002 &  0.395 $\pm$ 0.001 & 0.162 $\pm$ 0.003 &  0.212 $\pm$ 0.007 &  0.291 $\pm$ 0.007 &  0.157 $\pm$ 0.004 &  0.187 $\pm$ 0.002 &  0.205 $\pm$ 0.002 &  0.302 $\pm$ 0.233 &  0.178 $\pm$ 0.019 &  \textbf{0.136} $\pm$ 0.001 \\
            Email-Eu & 0.199 $\pm$ 0.006 &  0.164 $\pm$ 0.004 &  0.232 $\pm$ 0.009 &  0.291 $\pm$ 0.001 &  0.268 $\pm$ 0.002 &  0.154 $\pm$ 0.004 &  0.300 $\pm$ 0.004 &  OutOfMemory  &  0.158 $\pm$ 0.005 &  0.168 $\pm$ 0.015 &  \textbf{0.151} $\pm$ 0.009 \\ 
            Stack-Biology & 0.536 $\pm$ 0.001 &  0.467 $\pm$ 0.003 &  0.266 $\pm$ 0.007 &  0.202 $\pm$ 0.003 &  0.237 $\pm$ 0.016 &  0.235 $\pm$ 0.006 &  0.263 $\pm$ 0.004 &  0.311 $\pm$ 0.026 &  0.200 $\pm$ 0.002 &  0.259 $\pm$ 0.022 &  \textbf{0.152} $\pm$ 0.002 \\ 
            Stack-Physics & 0.532 $\pm$ 0.001 &  0.482 $\pm$ 0.001 &  0.286 $\pm$ 0.036 &  0.219 $\pm$ 0.008 &  0.289 $\pm$ 0.004 &  0.227 $\pm$ 0.006 &  0.285 $\pm$ 0.050 &  0.292 $\pm$ 0.005 &  0.162 $\pm$ 0.008 &  0.185 $\pm$ 0.021 &  \textbf{0.141} $\pm$ 0.003 \\ 
            \bottomrule
        \end{tabular}}
    \end{center}
\end{table*}

\begin{table*}[!ht]
    \vspace{-2mm}
    \caption{\label{tab:downstream}\underline{\smash{Real-World Applications of \method:}} In the three considered applications, utilizing hypergraphs with edge-dependent node labels predicted by \method (1) consistently outperforms using hypergraphs without such labels, (2) tends to outperform using edge-dependent node labels obtained by AST or HST (especially for clustering), and (3) performs comparably to and sometimes even better (especially for ranking aggregation) than using ground-truth labels.
    }
    \hspace{-1cm}
    \begin{subtable}{.35\linewidth}
        \centering
        \caption{Ranking Aggregation (Accuracy)}
        \scalebox{0.8}{
    	\begin{tabular}{l|c c}
    		\toprule
    		Method & Halo & H-Index \\ 
    		\midrule
    		RW~\cite{chitra2019random} w/ Ground Truth & \underline{0.711} & 0.675 \\
            \midrule
            RW~\cite{chitra2019random} w/ \method & \textbf{0.714} & 0.693 \\
            RW~\cite{chitra2019random} w/ HST & 0.707 & \underline{0.695} \\
            RW~\cite{chitra2019random} w/ AST & 0.706 & \textbf{0.696} \\
            \midrule
    		RW~\cite{chitra2019random} w/o Labels & 0.532 & 0.654 \\
    		\bottomrule
        \end{tabular}}
    \end{subtable}
    \hspace{-0.4cm}
    \begin{subtable}{.35\linewidth}
        \centering
        \caption{Clustering (NMI: The higher, the better)}
        \scalebox{0.8}{
        \begin{tabular}{l|c c}
			\toprule
			Method & DBLP & AMiner \\
			\midrule
			RDC-Spec~\cite{hayashi2020hypergraph} w/ GroundTruth & \textbf{0.221} & \textbf{0.359} \\
            \midrule
			RDC-Spec~\cite{hayashi2020hypergraph} w/ \method & \underline{0.184} & \underline{0.352} \\
            RDC-Spec~\cite{hayashi2020hypergraph} w/ HST & 0.166 & 0.339 \\
            RDC-Spec~\cite{hayashi2020hypergraph} w/ AST & 0.168 & 0.332 \\
            \midrule
			RDC-Spec~\cite{hayashi2020hypergraph} w/o Labels & 0.163 & 0.338 \\
			\bottomrule
		\end{tabular}}
	\end{subtable}
	\begin{subtable}{.3\linewidth}
	    \centering
	    \caption{Product Return Prediction (F1)}
        \scalebox{0.8}{
        \begin{tabular}{l|c c}
        	\toprule
        	Method & Synthetic E-tail \\
        	\midrule
        	HyperGO~\cite{li2018tail} w/ GroundTruth & \textbf{0.738}\\
            \midrule
        	HyperGO~\cite{li2018tail} w/ \method & 0.723 \\
        	HyperGO~\cite{li2018tail} w/ HST & \underline{0.724} \\
        	HyperGO~\cite{li2018tail} w/ AST & 0.721 \\
            \midrule
        	HyperGO~\cite{li2018tail} w/o Labels &  0.718 \\
        	\bottomrule
        \end{tabular}}
    \end{subtable}
    \vspace{2mm}
\end{table*}


\begin{table}[t]
    \vspace{-2mm}
    \begin{center}
        \caption{\label{tab:ablresult} \underline{\smash{Ablation Study of \method:}} The performance of \method is largely improved by \WithinAttention and \PE, as shown in the average rankings of models.}
        \scalebox{0.77}{
            \begin{tabular}{l l| c c c}
            \toprule
            Dataset & Metric & w/o \WithinAttention & w/o \PE & \method\\
            \midrule
            \multirow{2}{*}{\shortstack[l]{Coauth-\\DBLP}} & MicroF1 & 0.581 $\pm$ 0.004 & 0.591 $\pm$ 0.003 & \textbf{0.605} $\pm$ 0.002 \\
             & MacroF1 & 0.577 $\pm$ 0.003 & 0.584 $\pm$ 0.003 & \textbf{0.595} $\pm$ 0.002 \\
            \midrule
            \multirow{2}{*}{\shortstack[l]{Coauth-\\AMiner}} & MicroF1 & 0.604 $\pm$ 0.010 & 0.583 $\pm$ 0.095 & \textbf{0.630} $\pm$ 0.005 \\
             & MacroF1 & 0.592 $\pm$ 0.013 & 0.536 $\pm$ 0.174 & \textbf{0.623} $\pm$ 0.007 \\
            \midrule
            \multirow{2}{*}{\shortstack[l]{Email-\\Enron}} & MicroF1 & 0.812 $\pm$ 0.008 & 0.825 $\pm$ 0.001 & \textbf{0.826} $\pm$ 0.001 \\
             & MacroF1 & 0.747 $\pm$ 0.014 & \textbf{0.762} $\pm$ 0.004 & 0.760 $\pm$ 0.004 \\
            \midrule
            \multirow{2}{*}{\shortstack[l]{Email-\\Eu}} & MicroF1 & 0.651 $\pm$ 0.019 & 0.670 $\pm$ 0.000 & \textbf{0.671} $\pm$ 0.000 \\
             & MacroF1 & 0.630 $\pm$ 0.018 & 0.638 $\pm$ 0.002 & \textbf{0.646} $\pm$ 0.003 \\
            \midrule
            \multirow{2}{*}{\shortstack[l]{Stack-\\Biology}} & MicroF1 & 0.723 $\pm$ 0.002 & 0.732 $\pm$ 0.002 & \textbf{0.742} $\pm$ 0.003 \\
             & MacroF1 & 0.656 $\pm$ 0.005 & 0.672 $\pm$ 0.004 & \textbf{0.686} $\pm$ 0.004 \\
            \midrule
            \multirow{2}{*}{\shortstack[l]{Stack-\\Physics}} & MicroF1 & 0.752 $\pm$ 0.005 & 0.765 $\pm$ 0.002 & \textbf{0.770} $\pm$ 0.003 \\
             & MacroF1 & 0.675 $\pm$ 0.010 & 0.688 $\pm$ 0.008 & \textbf{0.707} $\pm$ 0.004 \\
            \midrule
            \midrule
            \multirow{2}{*}{\shortstack[l]{AVG.\\Ranking}} & MicroF1 & 2.83 & 2.17 & \textbf{1.00} \\
            & MacroF1 & 2.83 & 2.00 & \textbf{1.17} \\
            \bottomrule
        \end{tabular}}
    \end{center}
\end{table}

\subsection{Q1. Edge-Dependent Node Classification}

To evaluate the performance of \method for the edge-dependent node classification, we measure the predictive performances 
using Micro-F1 and Macro-F1 scores on test data, averaged over five runs. 
Table~\ref{tab:expresult} shows that \method \textit{consistently} achieves the highest scores in terms of both Micro-F1 and Macro-F1. 
While HST performs well overall and ranks second, the statistical analysis using the Wilcoxon signed-rank test indicates that \method is significantly better than HST with a p-value less than 0.05 in almost all cases (only except for Macro-F1 in Email-Enron and Micro-F1 in Email-Eu). 
Some other models, including HGNN and HAT, perform well on specific datasets but fail to achieve overall success. 
HNN runs out of memory on the email-Eu dataset due to the large size of the random-walk hyperedge transition matrix.\footnote{Implementation details of HNN can be found in~\cite{github}.}

We would like to emphasize that HST, AST, and \method utilize different aggregation methods, and relationships among nodes within each hyperedge (or among hyperedges incident to each node) are explicitly considered only in HST and \method. Additionally, positional encoding is used only in \method (see Section~\ref{method:comp}).
Thus, the fact that AST underperforms HST and \method supports the effectiveness of \WithinAttention, and the fact that \method outperforms HST supports the usefulness of \PE and our aggregation method.

\begin{figure}[!t]
    \vspace{-3mm}
    \begin{minipage}{0.74\linewidth}
        \includegraphics[width=1.03\linewidth]{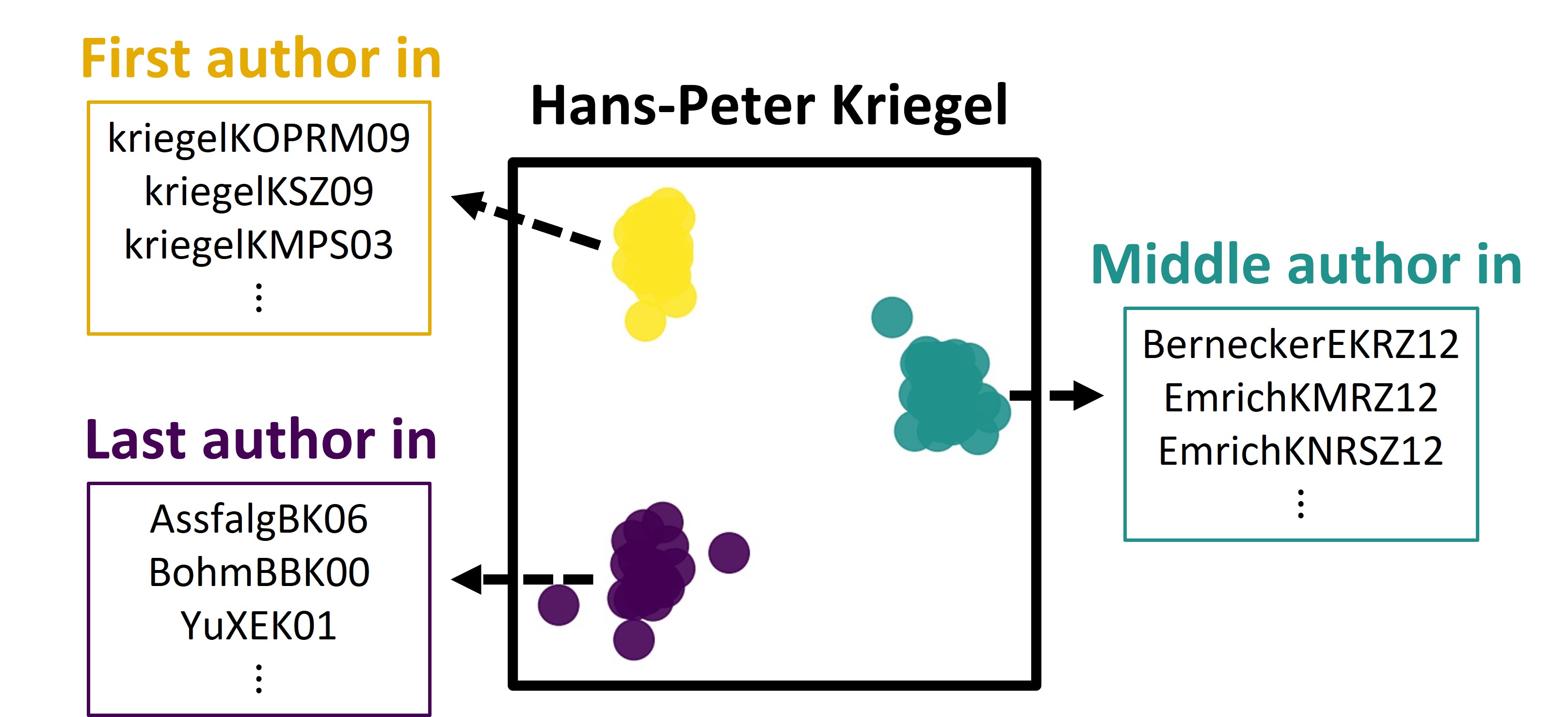}
    \end{minipage}
    \begin{minipage}{0.25\linewidth}
        \includegraphics[width=1.0\linewidth]{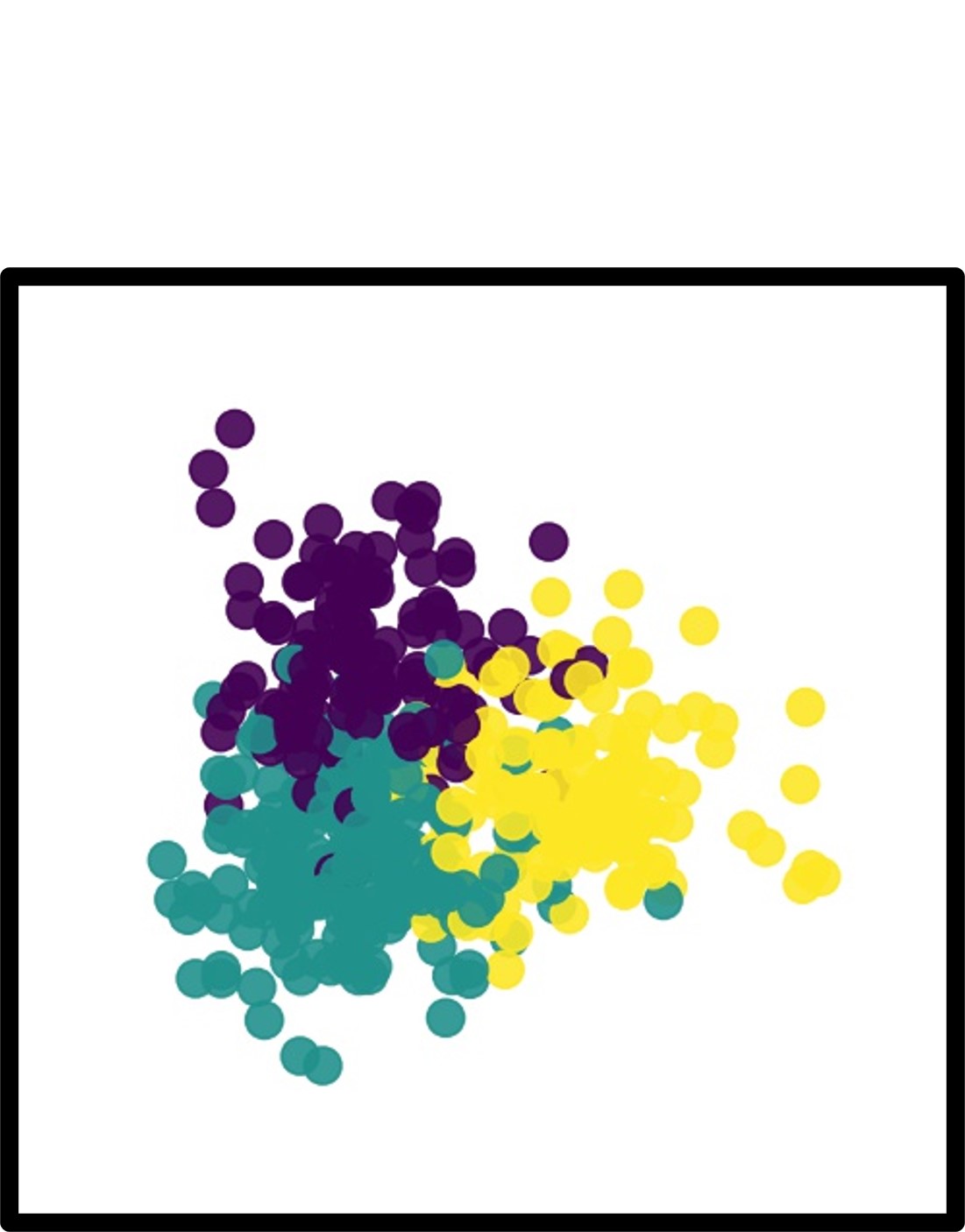}
    \end{minipage}
    
    \begin{minipage}{0.56\linewidth} \small
        (a) Embeddings of publications containing "Hans-Peter Kriegel"
    \end{minipage}
    \quad
    \quad
    \begin{minipage}{0.32\linewidth} \small
        (b) All concatenated embeddings of incident hyperedges and nodes
    \end{minipage}
    
    \caption{\label{fig:emb_vis}\underline{\smash{Visualization of embeddings from \method.}} (a)  The embeddings of all publications of ``Hans-Peter Kriegel'' visualized using LDA. (b) The concatenated embeddings of all incident pairs of nodes and hyperedges in the test set, visualized in the same manner.  
    Note that they are clearly clustered based on the edge-dependent labels.} 
\end{figure}

Figure~\ref{fig:emb_vis} visually demonstrates the effectiveness of \method using the Coauthorship-DBLP dataset. The visualization showcases (a) the embeddings of hyperedges containing a specific node and (b) the concatenated embeddings of all incident pairs of nodes and hyperedges in the test set. Notably, the embeddings exhibit clear distinctions based on the edge-dependent labels.

\subsection{Q2. Node Label Distribution Preservation}\label{sec:evaluation:labeldist}

In the task of edge-dependent node classification, each node may have different labels for different edges. That is, each node has its own ground-truth label distribution (node-level label distribution), which describes how many times it has each label. 
In this experiment, we aim to check how well \method preserves such node-level label distributions compared to the other baseline approaches.
To this end, we measure the Jensen-Shannon divergence (JSD) \cite{lin1991divergence} between the ground-truth and the predicted node-level label distribution for each node in the test data.
The average JSD over all nodes is used as a metric to evaluate how well the model captures the ground-truth node-level label distributions. As shown in Table~\ref{tab:jsdresult}, 
\method outperforms all other models, with the lowest JSD in all datasets. 
This result suggests that \method preserves well the ground-truth node-level label distributions.



\subsection{Q3. Ablation Study of \method}

For an ablation study, we compare the performance of \method with two variants of it in Table~\ref{tab:ablresult}:
(a) \textit{\method w/o \WithinAttention}, where \WithinAttention is removed and only the aggregation is used for message passing, and (b) \textit{\method w/o \PE}, which computes \WithinAttention without positional encodings. 

\begin{figure}[!t]
    \vspace{-3mm}
    \begin{minipage}{0.46\linewidth}
        \includegraphics[width=1.03\linewidth]{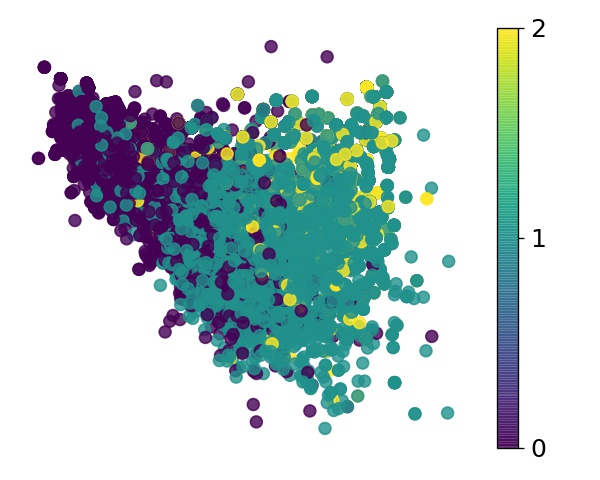}
    \end{minipage}
    \begin{minipage}{0.46\linewidth}
        \includegraphics[width=1.0\linewidth]{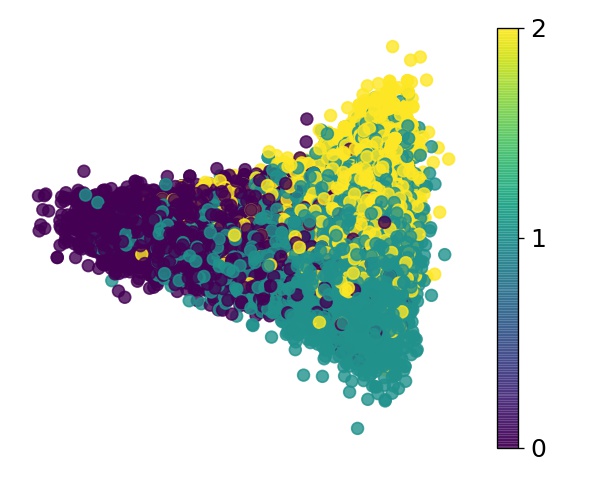}
    \end{minipage}
    
    \begin{minipage}{0.45\linewidth} \small
        (a) Before \WithinAttention
    \end{minipage}
    \quad
    \begin{minipage}{0.35\linewidth} \small
        (b) After \WithinAttention
    \end{minipage}
    
    \caption{\label{fig:att_vis}\underline{\smash{Visualization of the effect of \WithinAttention.}} Node embeddings in StackOverflow-Biology (a) before and (b) after the \WithinAttention module are visualized using LDA.
    \WithinAttention makes node embeddings better distinguished based on edge-dependent labels, represented by different colors.}
\end{figure}

\smallsection{Importance of \WithinAttention}
As shown in Table~\ref{tab:ablresult}, \method consistently outperforms the variant without \WithinAttention. 
This result demonstrates the importance of \WithinAttention in achieving accurate edge-dependent node label classification. That is, considering relationships among nodes within each hyperedge is crucial for precise classification.
In Appendix~\ref{appendix:abl_partial}, we also empirically confirm that the improvement is also contributed by \WithinAttention among hyperedges containing each node, by using a variant without it.
Figure~\ref{fig:att_vis} visually illustrates the effectiveness of \WithinAttention by demonstrating \WithinAttention makes node embeddings better distinguished based on edge-dependent labels.

\smallsection{Usefulness of \PE} Table~\ref{tab:ablresult} also indicates that \method also performs better than the variant without \PE in terms of both Micro F1 and Macro F1 scores across most datasets. 
This result supports that \PE contributes to the improvement in performance. 
However, in the Email-Enron dataset, \PE does not have a significant impact, which aligns with the lowest correlation between node labels and node centrality orders in Table~\ref{tab:datasets}.
In Appendix~\ref{appendix:abl_pe}, we show that \PE outperforms several positional encoding schemes, including global node-centrality ranks.




\subsection{Q4. Usefulness in Downstream Tasks}

To evaluate the usefulness of \method, we conduct experiments on three downstream tasks: ranking aggregation, clustering, and product return prediction (see Section~\ref{benchmark_application}). We compare the performance of the tasks using three different types of inputs: hypergraphs with (a) ground-truth edge-dependent node labels, (b) labels predicted by hypergraph neural networks (spec., \method, HST, and AST) trained for our problem (i.e., edge-dependent node classification problem), and (c) no labels, respectively.
We aim to demonstrate that even imperfect edge-dependent node labels predicted by the trained model (especially, by \method) lead to better performances, compared to those without labels. 
Refer to \cite{github} for details of the edge-dependent node labels used in each task.

\smallsection{Ranking Aggregation}
This task aims to correctly predict the global node ranks using local node ranks in subsets (hyperedges). 
We use a recent random-walk-based method~\cite{chitra2019random}, where a random walker selects a node in each hyperedge proportionally to its rank within it, and the global node ranking is determined by the stationary distribution.
We use two real-world datasets: (1) the Halo 2 game dataset, where scores (edge-dependent node labels) of players (nodes) in matches (hyperedges) of up to 8 players are given, and the global rankings of players need to be inferred; (2) the AMiner dataset, where the order of authorship (i.e., edge-dependent labels) in each publication is given to predict the rankings of all authors in terms of their H-index.
For evaluation, we measure the accuracy in identifying the node with a higher rank for each node pair. 

\smallsection{Clustering}
From two co-authorship datasets, (1) DBLP and (2) AMiner, we group publications by venues as their ground-truth clusters. 
We use RDC-Spec~\cite{hayashi2020hypergraph}, a well-performing method that uses spectral clustering and weighs authors differently according to their order (edge-dependent label) in each publication, as the backbone hypergraph clustering algorithm. For evaluation, we measure Normalized mutual information (NMI).

\smallsection{Product Return Prediction}
We use a state-of-the-art method, HyperGo ~\cite{li2018tail}, as the predictor for the returned product.
It utilizes the information about the count of products in each basket, which corresponds to edge-dependent labels.
We evaluate by F1 score of the predicted return probability of products in each target basket.

\smallsection{Results}
As shown in Table~\ref{tab:downstream}, across all the examined applications, the utilization of imperfect edge-dependent node labels predicted by \method consistently leads to performance improvements compared to the results obtained without labels.
Furthermore, these enhancements generally surpass those achieved by HST and AST.
Surprisingly, in ranking aggregation, the performance is even higher than what is achieved by using ground-truth labels. 
We would like to emphasize that \method can be applied to any algorithm that utilizes edge-dependent labels, and \method is not specialized to any of the methods considered for the tasks.
The three application tasks are described in greater detail in \cite{github}.

	\section{Conclusions}
	\label{sec:conclusion}
	In this work, we propose the edge-dependent node classification problem, which is a new benchmark task for hypergraph neural networks with various real-life applications. In order to tackle this problem, we devise \method, a hypergraph neural network for considering edge-dependent relationships between node pairs within each hyperedge.
In our experiments with $6$ real-world hypergraphs,  \method consistently outperformed 10 competitors.
Our contributions are summarized as follows:
\begin{itemize}[leftmargin=*]
    \item \textbf{New Problem}: We formulate the edge-dependent node classification problem with six real-world datasets and three applications. 
    \item \textbf{Effective Model}: We propose \method, a novel hypergraph neural network equipped with the \WithinAttention module and \PE. They are designed to capture edge-dependent relationships between node pairs within each hyperedge.
    \item \textbf{Extensive Experiment}: We empirically show the advantage of \method over $10$ competitors and the effectiveness of each component of it.
    We also demonstrated the usefulness of \method in three applications.%
    
\end{itemize}
For reproducibility, we make the code and data available at \cite{github}.

    {\small \smallsection{Acknowledgements} This work was supported by National Research Foundation of Korea (NRF) grant funded by the Korea government (MSIT) (No. NRF-2020R1C1C1008296) and Institute of Information \& Communications Technology Planning \& Evaluation (IITP) grant funded by the Korea government (MSIT) (No. 2022-0-00157, Robust, Fair, Extensible Data-Centric Continual Learning) (No. 2019-0-00075, Artificial Intelligence Graduate School Program (KAIST)).
    }

    \bibliographystyle{ACM-Reference-Format}
    \balance
    \bibliography{BIB/ref}


\begin{thebibliography}{68}


\ifx \showCODEN    \undefined \def \showCODEN     #1{\unskip}     \fi
\ifx \showDOI      \undefined \def \showDOI       #1{#1}\fi
\ifx \showISBNx    \undefined \def \showISBNx     #1{\unskip}     \fi
\ifx \showISBNxiii \undefined \def \showISBNxiii  #1{\unskip}     \fi
\ifx \showISSN     \undefined \def \showISSN      #1{\unskip}     \fi
\ifx \showLCCN     \undefined \def \showLCCN      #1{\unskip}     \fi
\ifx \shownote     \undefined \def \shownote      #1{#1}          \fi
\ifx \showarticletitle \undefined \def \showarticletitle #1{#1}   \fi
\ifx \showURL      \undefined \def \showURL       {\relax}        \fi
\providecommand\bibfield[2]{#2}
\providecommand\bibinfo[2]{#2}
\providecommand\natexlab[1]{#1}
\providecommand\showeprint[2][]{arXiv:#2}

\bibitem[\protect\citeauthoryear{Abboud, Ceylan, Grohe, and Lukasiewicz}{Abboud
  et~al\mbox{.}}{2020}]%
        {abboud2020surprising}
\bibfield{author}{\bibinfo{person}{Ralph Abboud}, \bibinfo{person}{Ismail~Ilkan
  Ceylan}, \bibinfo{person}{Martin Grohe}, {and} \bibinfo{person}{Thomas
  Lukasiewicz}.} \bibinfo{year}{2020}\natexlab{}.
\newblock \showarticletitle{The surprising power of graph neural networks with
  random node initialization}.
\newblock \bibinfo{journal}{\emph{arXiv preprint arXiv:2010.01179}}
  (\bibinfo{year}{2020}).
\newblock


\bibitem[\protect\citeauthoryear{Ahmadi}{Ahmadi}{2020}]%
        {ahmadi2020memory}
\bibfield{author}{\bibinfo{person}{Amir Hosein~Khas Ahmadi}.}
  \bibinfo{year}{2020}\natexlab{}.
\newblock \emph{\bibinfo{title}{Memory-based graph networks}}.
\newblock \bibinfo{thesistype}{Ph.D. Dissertation}. \bibinfo{school}{University
  of Toronto (Canada)}.
\newblock


\bibitem[\protect\citeauthoryear{Aponte, Rossi, Guo, Hoffswell, Lipka, Xiao,
  Chan, Koh, and Ahmed}{Aponte et~al\mbox{.}}{2022}]%
        {aponte2022hypergraph}
\bibfield{author}{\bibinfo{person}{Ryan Aponte}, \bibinfo{person}{Ryan~A
  Rossi}, \bibinfo{person}{Shunan Guo}, \bibinfo{person}{Jane Hoffswell},
  \bibinfo{person}{Nedim Lipka}, \bibinfo{person}{Chang Xiao},
  \bibinfo{person}{Gromit Chan}, \bibinfo{person}{Eunyee Koh}, {and}
  \bibinfo{person}{Nesreen Ahmed}.} \bibinfo{year}{2022}\natexlab{}.
\newblock \showarticletitle{A Hypergraph Neural Network Framework for Learning
  Hyperedge-Dependent Node Embeddings}.
\newblock \bibinfo{journal}{\emph{arXiv preprint arXiv:2212.14077}}
  (\bibinfo{year}{2022}).
\newblock


\bibitem[\protect\citeauthoryear{Ba, Kiros, and Hinton}{Ba
  et~al\mbox{.}}{2016}]%
        {ba2016layer}
\bibfield{author}{\bibinfo{person}{Jimmy~Lei Ba}, \bibinfo{person}{Jamie~Ryan
  Kiros}, {and} \bibinfo{person}{Geoffrey~E Hinton}.}
  \bibinfo{year}{2016}\natexlab{}.
\newblock \showarticletitle{Layer normalization}.
\newblock \bibinfo{journal}{\emph{arXiv:1607.06450}} (\bibinfo{year}{2016}).
\newblock


\bibitem[\protect\citeauthoryear{Bai, Zhang, and Torr}{Bai
  et~al\mbox{.}}{2021}]%
        {bai2021hypergraph}
\bibfield{author}{\bibinfo{person}{Song Bai}, \bibinfo{person}{Feihu Zhang},
  {and} \bibinfo{person}{Philip~HS Torr}.} \bibinfo{year}{2021}\natexlab{}.
\newblock \showarticletitle{Hypergraph convolution and hypergraph attention}.
\newblock \bibinfo{journal}{\emph{Pattern Recognition}}  \bibinfo{volume}{110}
  (\bibinfo{year}{2021}), \bibinfo{pages}{107637}.
\newblock


\bibitem[\protect\citeauthoryear{Belkin and Niyogi}{Belkin and Niyogi}{2003}]%
        {belkin2003laplacian}
\bibfield{author}{\bibinfo{person}{Mikhail Belkin} {and}
  \bibinfo{person}{Partha Niyogi}.} \bibinfo{year}{2003}\natexlab{}.
\newblock \showarticletitle{Laplacian eigenmaps for dimensionality reduction
  and data representation}.
\newblock \bibinfo{journal}{\emph{Neural computation}} \bibinfo{volume}{15},
  \bibinfo{number}{6} (\bibinfo{year}{2003}), \bibinfo{pages}{1373--1396}.
\newblock


\bibitem[\protect\citeauthoryear{Benson, Abebe, Schaub, Jadbabaie, and
  Kleinberg}{Benson et~al\mbox{.}}{2018a}]%
        {benson2018simplicial}
\bibfield{author}{\bibinfo{person}{Austin~R Benson}, \bibinfo{person}{Rediet
  Abebe}, \bibinfo{person}{Michael~T Schaub}, \bibinfo{person}{Ali Jadbabaie},
  {and} \bibinfo{person}{Jon Kleinberg}.} \bibinfo{year}{2018}\natexlab{a}.
\newblock \showarticletitle{Simplicial closure and higher-order link
  prediction}.
\newblock \bibinfo{journal}{\emph{Proceedings of the National Academy of
  Sciences}} \bibinfo{volume}{115}, \bibinfo{number}{48}
  (\bibinfo{year}{2018}), \bibinfo{pages}{E11221--E11230}.
\newblock


\bibitem[\protect\citeauthoryear{Benson, Kumar, and Tomkins}{Benson
  et~al\mbox{.}}{2018b}]%
        {benson2018sequences}
\bibfield{author}{\bibinfo{person}{Austin~R Benson}, \bibinfo{person}{Ravi
  Kumar}, {and} \bibinfo{person}{Andrew Tomkins}.}
  \bibinfo{year}{2018}\natexlab{b}.
\newblock \showarticletitle{Sequences of sets}. In
  \bibinfo{booktitle}{\emph{KDD}}.
\newblock


\bibitem[\protect\citeauthoryear{Bonacich}{Bonacich}{1987}]%
        {bonacich1987power}
\bibfield{author}{\bibinfo{person}{Phillip Bonacich}.}
  \bibinfo{year}{1987}\natexlab{}.
\newblock \showarticletitle{Power and centrality: A family of measures}.
\newblock \bibinfo{journal}{\emph{AJS}} \bibinfo{volume}{92},
  \bibinfo{number}{5} (\bibinfo{year}{1987}), \bibinfo{pages}{1170--1182}.
\newblock


\bibitem[\protect\citeauthoryear{Bu, Lee, and Shin}{Bu et~al\mbox{.}}{2023}]%
        {bu2023hypercore}
\bibfield{author}{\bibinfo{person}{Fanchen Bu}, \bibinfo{person}{Geon Lee},
  {and} \bibinfo{person}{Kijung Shin}.} \bibinfo{year}{2023}\natexlab{}.
\newblock \showarticletitle{Hypercore Decomposition for Non-Fragile Hyperedges:
  Concepts, Algorithms, Observations, and Applications}.
\newblock \bibinfo{journal}{\emph{arXiv preprint arXiv:2301.08440}}
  (\bibinfo{year}{2023}).
\newblock


\bibitem[\protect\citeauthoryear{Chen, Wei, Huang, Ding, and Li}{Chen
  et~al\mbox{.}}{2020}]%
        {chen2020simple}
\bibfield{author}{\bibinfo{person}{Ming Chen}, \bibinfo{person}{Zhewei Wei},
  \bibinfo{person}{Zengfeng Huang}, \bibinfo{person}{Bolin Ding}, {and}
  \bibinfo{person}{Yaliang Li}.} \bibinfo{year}{2020}\natexlab{}.
\newblock \showarticletitle{Simple and deep graph convolutional networks}. In
  \bibinfo{booktitle}{\emph{ICML}}.
\newblock


\bibitem[\protect\citeauthoryear{Chien, Pan, Peng, and Milenkovic}{Chien
  et~al\mbox{.}}{2022}]%
        {chien2021you}
\bibfield{author}{\bibinfo{person}{Eli Chien}, \bibinfo{person}{Chao Pan},
  \bibinfo{person}{Jianhao Peng}, {and} \bibinfo{person}{Olgica Milenkovic}.}
  \bibinfo{year}{2022}\natexlab{}.
\newblock \showarticletitle{You are allset: A multiset function framework for
  hypergraph neural networks}. In \bibinfo{booktitle}{\emph{ICLR}}.
\newblock


\bibitem[\protect\citeauthoryear{Chitra and Raphael}{Chitra and
  Raphael}{2019}]%
        {chitra2019random}
\bibfield{author}{\bibinfo{person}{Uthsav Chitra} {and}
  \bibinfo{person}{Benjamin Raphael}.} \bibinfo{year}{2019}\natexlab{}.
\newblock \showarticletitle{Random walks on hypergraphs with edge-dependent
  vertex weights}. In \bibinfo{booktitle}{\emph{ICML}}.
\newblock


\bibitem[\protect\citeauthoryear{Choe, Kim, Yoo, and Shin}{Choe
  et~al\mbox{.}}{2023}]%
        {github}
\bibfield{author}{\bibinfo{person}{Minyoung Choe}, \bibinfo{person}{Sunwoo
  Kim}, \bibinfo{person}{Jaemin Yoo}, {and} \bibinfo{person}{Kijung Shin}.}
  \bibinfo{year}{2023}\natexlab{}.
\newblock \bibinfo{booktitle}{\emph{Classification of Edge-dependent Labels of
  Nodes in Hypergraphs (Code, Datasets, and Online Appendix)}}.
\newblock
\urldef\tempurl%
\url{https://github.com/young917/EdgeDependentNodeLabel}
\showURL{%
\tempurl}


\bibitem[\protect\citeauthoryear{Cramer}{Cramer}{1946}]%
        {cramer1946mathematical}
\bibfield{author}{\bibinfo{person}{Harold Cramer}.}
  \bibinfo{year}{1946}\natexlab{}.
\newblock \showarticletitle{Mathematical methods of statistics, Princeton
  Univ}.
\newblock \bibinfo{journal}{\emph{PUP}} (\bibinfo{year}{1946}).
\newblock


\bibitem[\protect\citeauthoryear{Do, Yoon, Hooi, and Shin}{Do
  et~al\mbox{.}}{2020}]%
        {do2020structural}
\bibfield{author}{\bibinfo{person}{Manh~Tuan Do}, \bibinfo{person}{Se-eun
  Yoon}, \bibinfo{person}{Bryan Hooi}, {and} \bibinfo{person}{Kijung Shin}.}
  \bibinfo{year}{2020}\natexlab{}.
\newblock \showarticletitle{Structural patterns and generative models of
  real-world hypergraphs}. In \bibinfo{booktitle}{\emph{KDD}}.
\newblock


\bibitem[\protect\citeauthoryear{Dong, Sawin, and Bengio}{Dong
  et~al\mbox{.}}{2020}]%
        {dong2020hnhn}
\bibfield{author}{\bibinfo{person}{Yihe Dong}, \bibinfo{person}{Will Sawin},
  {and} \bibinfo{person}{Yoshua Bengio}.} \bibinfo{year}{2020}\natexlab{}.
\newblock \showarticletitle{HNHN: hypergraph networks with hyperedge neurons}.
  In \bibinfo{booktitle}{\emph{ICML Graph Representation Learning and Beyond
  Workshop}}.
\newblock


\bibitem[\protect\citeauthoryear{Dwivedi and Bresson}{Dwivedi and
  Bresson}{2021}]%
        {dwivedi2020generalization}
\bibfield{author}{\bibinfo{person}{Vijay~Prakash Dwivedi} {and}
  \bibinfo{person}{Xavier Bresson}.} \bibinfo{year}{2021}\natexlab{}.
\newblock \showarticletitle{A generalization of transformer networks to
  graphs}. In \bibinfo{booktitle}{\emph{AAAI Workshop on Deep Learning on
  Graphs: Methods and Applications}}.
\newblock


\bibitem[\protect\citeauthoryear{Dwivedi, Luu, Laurent, Bengio, and
  Bresson}{Dwivedi et~al\mbox{.}}{2022}]%
        {dwivedi2021graph}
\bibfield{author}{\bibinfo{person}{Vijay~Prakash Dwivedi},
  \bibinfo{person}{Anh~Tuan Luu}, \bibinfo{person}{Thomas Laurent},
  \bibinfo{person}{Yoshua Bengio}, {and} \bibinfo{person}{Xavier Bresson}.}
  \bibinfo{year}{2022}\natexlab{}.
\newblock \showarticletitle{Graph neural networks with learnable structural and
  positional representations}. In \bibinfo{booktitle}{\emph{ICLR}}.
\newblock


\bibitem[\protect\citeauthoryear{Feng, You, Zhang, Ji, and Gao}{Feng
  et~al\mbox{.}}{2019}]%
        {feng2019hypergraph}
\bibfield{author}{\bibinfo{person}{Yifan Feng}, \bibinfo{person}{Haoxuan You},
  \bibinfo{person}{Zizhao Zhang}, \bibinfo{person}{Rongrong Ji}, {and}
  \bibinfo{person}{Yue Gao}.} \bibinfo{year}{2019}\natexlab{}.
\newblock \showarticletitle{Hypergraph neural networks}. In
  \bibinfo{booktitle}{\emph{AAAI}}.
\newblock


\bibitem[\protect\citeauthoryear{Gu, Chen, and Krenn}{Gu et~al\mbox{.}}{2020}]%
        {gu2020quantum}
\bibfield{author}{\bibinfo{person}{Xuemei Gu}, \bibinfo{person}{Lijun Chen},
  {and} \bibinfo{person}{Mario Krenn}.} \bibinfo{year}{2020}\natexlab{}.
\newblock \showarticletitle{Quantum experiments and hypergraphs: Multiphoton
  sources for quantum interference, quantum computation, and quantum
  entanglement}.
\newblock \bibinfo{journal}{\emph{PRA}} \bibinfo{volume}{101},
  \bibinfo{number}{3} (\bibinfo{year}{2020}), \bibinfo{pages}{033816}.
\newblock


\bibitem[\protect\citeauthoryear{Hada, Shevade, et~al\mbox{.}}{Hada
  et~al\mbox{.}}{2021}]%
        {hada2021hypertenet}
\bibfield{author}{\bibinfo{person}{Deepesh Hada}, \bibinfo{person}{Shirish
  Shevade}, {et~al\mbox{.}}} \bibinfo{year}{2021}\natexlab{}.
\newblock \showarticletitle{HyperTeNet: Hypergraph and Transformer-based Neural
  Network for Personalized List Continuation}. In
  \bibinfo{booktitle}{\emph{ICDM}}.
\newblock


\bibitem[\protect\citeauthoryear{Hayashi, Aksoy, Park, and Park}{Hayashi
  et~al\mbox{.}}{2020}]%
        {hayashi2020hypergraph}
\bibfield{author}{\bibinfo{person}{Koby Hayashi}, \bibinfo{person}{Sinan~G
  Aksoy}, \bibinfo{person}{Cheong~Hee Park}, {and} \bibinfo{person}{Haesun
  Park}.} \bibinfo{year}{2020}\natexlab{}.
\newblock \showarticletitle{Hypergraph random walks, laplacians, and
  clustering}. In \bibinfo{booktitle}{\emph{CIKM}}.
\newblock


\bibitem[\protect\citeauthoryear{He, Zhang, Ren, and Sun}{He
  et~al\mbox{.}}{2015}]%
        {he2015delving}
\bibfield{author}{\bibinfo{person}{Kaiming He}, \bibinfo{person}{Xiangyu
  Zhang}, \bibinfo{person}{Shaoqing Ren}, {and} \bibinfo{person}{Jian Sun}.}
  \bibinfo{year}{2015}\natexlab{}.
\newblock \showarticletitle{Delving deep into rectifiers: Surpassing
  human-level performance on imagenet classification}. In
  \bibinfo{booktitle}{\emph{ICCV}}.
\newblock


\bibitem[\protect\citeauthoryear{He, Zhang, Ren, and Sun}{He
  et~al\mbox{.}}{2016}]%
        {he2016deep}
\bibfield{author}{\bibinfo{person}{Kaiming He}, \bibinfo{person}{Xiangyu
  Zhang}, \bibinfo{person}{Shaoqing Ren}, {and} \bibinfo{person}{Jian Sun}.}
  \bibinfo{year}{2016}\natexlab{}.
\newblock \showarticletitle{Deep residual learning for image recognition}. In
  \bibinfo{booktitle}{\emph{CVPR}}.
\newblock


\bibitem[\protect\citeauthoryear{Huang and Yang}{Huang and Yang}{2021}]%
        {huang2021unignn}
\bibfield{author}{\bibinfo{person}{Jing Huang} {and} \bibinfo{person}{Jie
  Yang}.} \bibinfo{year}{2021}\natexlab{}.
\newblock \showarticletitle{Unignn: a unified framework for graph and
  hypergraph neural networks}. In \bibinfo{booktitle}{\emph{IJCAI}}.
\newblock


\bibitem[\protect\citeauthoryear{Hwang, Lee, and Shin}{Hwang
  et~al\mbox{.}}{2021}]%
        {hwang2021hyfer}
\bibfield{author}{\bibinfo{person}{Hyunjin Hwang}, \bibinfo{person}{Seungwoo
  Lee}, {and} \bibinfo{person}{Kijung Shin}.} \bibinfo{year}{2021}\natexlab{}.
\newblock \showarticletitle{HyFER: A Framework for Making Hypergraph Learning
  Easy, Scalable and Benchmarkable}. In \bibinfo{booktitle}{\emph{WWW Workshop
  on Graph Learning Benchmarks}}.
\newblock


\bibitem[\protect\citeauthoryear{Jiang, Mitzenmacher, and Thaler}{Jiang
  et~al\mbox{.}}{2017}]%
        {jiang2017parallel}
\bibfield{author}{\bibinfo{person}{Jiayang Jiang}, \bibinfo{person}{Michael
  Mitzenmacher}, {and} \bibinfo{person}{Justin Thaler}.}
  \bibinfo{year}{2017}\natexlab{}.
\newblock \showarticletitle{Parallel peeling algorithms}.
\newblock \bibinfo{journal}{\emph{TOPC}} \bibinfo{volume}{3},
  \bibinfo{number}{1} (\bibinfo{year}{2017}), \bibinfo{pages}{1--27}.
\newblock


\bibitem[\protect\citeauthoryear{Kim, Oh, and Hong}{Kim et~al\mbox{.}}{2021}]%
        {kim2021transformers}
\bibfield{author}{\bibinfo{person}{Jinwoo Kim}, \bibinfo{person}{Saeyoon Oh},
  {and} \bibinfo{person}{Seunghoon Hong}.} \bibinfo{year}{2021}\natexlab{}.
\newblock \showarticletitle{Transformers Generalize DeepSets and Can be
  Extended to Graphs \& Hypergraphs}. In \bibinfo{booktitle}{\emph{NeurIPS}}.
\newblock


\bibitem[\protect\citeauthoryear{Kingma and Ba}{Kingma and Ba}{2015}]%
        {kingma2014adam}
\bibfield{author}{\bibinfo{person}{Diederik~P Kingma} {and}
  \bibinfo{person}{Jimmy Ba}.} \bibinfo{year}{2015}\natexlab{}.
\newblock \showarticletitle{Adam: A method for stochastic optimization}. In
  \bibinfo{booktitle}{\emph{ICLR}}.
\newblock


\bibitem[\protect\citeauthoryear{Kondor and Vert}{Kondor and Vert}{2004}]%
        {kondor2004diffusion}
\bibfield{author}{\bibinfo{person}{Risi Kondor} {and}
  \bibinfo{person}{Jean-Philippe Vert}.} \bibinfo{year}{2004}\natexlab{}.
\newblock \showarticletitle{Diffusion kernels}.
\newblock \bibinfo{journal}{\emph{kernel methods in computational biology}}
  (\bibinfo{year}{2004}), \bibinfo{pages}{171--192}.
\newblock


\bibitem[\protect\citeauthoryear{Kovalenko, Romance, Vasilyeva, Aleja, Criado,
  Musatov, Raigorodskii, Flores, Samoylenko, Alfaro-Bittner,
  et~al\mbox{.}}{Kovalenko et~al\mbox{.}}{2022}]%
        {kovalenko2022vector}
\bibfield{author}{\bibinfo{person}{Kirill Kovalenko}, \bibinfo{person}{Miguel
  Romance}, \bibinfo{person}{Ekaterina Vasilyeva}, \bibinfo{person}{David
  Aleja}, \bibinfo{person}{Regino Criado}, \bibinfo{person}{Daniil Musatov},
  \bibinfo{person}{Andrei~M Raigorodskii}, \bibinfo{person}{Julio Flores},
  \bibinfo{person}{Ivan Samoylenko}, \bibinfo{person}{Karin Alfaro-Bittner},
  {et~al\mbox{.}}} \bibinfo{year}{2022}\natexlab{}.
\newblock \showarticletitle{Vector centrality in hypergraphs}.
\newblock \bibinfo{journal}{\emph{Chaos, Solitons \& Fractals}}
  \bibinfo{volume}{162} (\bibinfo{year}{2022}), \bibinfo{pages}{112397}.
\newblock


\bibitem[\protect\citeauthoryear{Kreuzer, Beaini, Hamilton, L{\'e}tourneau, and
  Tossou}{Kreuzer et~al\mbox{.}}{2021}]%
        {kreuzer2021rethinking}
\bibfield{author}{\bibinfo{person}{Devin Kreuzer}, \bibinfo{person}{Dominique
  Beaini}, \bibinfo{person}{Will Hamilton}, \bibinfo{person}{Vincent
  L{\'e}tourneau}, {and} \bibinfo{person}{Prudencio Tossou}.}
  \bibinfo{year}{2021}\natexlab{}.
\newblock \showarticletitle{Rethinking graph transformers with spectral
  attention}. In \bibinfo{booktitle}{\emph{NeurIPS}}.
\newblock


\bibitem[\protect\citeauthoryear{Lee, Choe, and Shin}{Lee
  et~al\mbox{.}}{2021}]%
        {lee2021hyperedges}
\bibfield{author}{\bibinfo{person}{Geon Lee}, \bibinfo{person}{Minyoung Choe},
  {and} \bibinfo{person}{Kijung Shin}.} \bibinfo{year}{2021}\natexlab{}.
\newblock \showarticletitle{How do hyperedges overlap in real-world
  hypergraphs?-patterns, measures, and generators}. In
  \bibinfo{booktitle}{\emph{WWW}}.
\newblock


\bibitem[\protect\citeauthoryear{Lee, Choe, and Shin}{Lee
  et~al\mbox{.}}{2022}]%
        {lee2022hashnwalk}
\bibfield{author}{\bibinfo{person}{Geon Lee}, \bibinfo{person}{Minyoung Choe},
  {and} \bibinfo{person}{Kijung Shin}.} \bibinfo{year}{2022}\natexlab{}.
\newblock \showarticletitle{HashNWalk: Hash and Random Walk Based Anomaly
  Detection in Hyperedge Streams}. In \bibinfo{booktitle}{\emph{IJCAI}}.
\newblock


\bibitem[\protect\citeauthoryear{Lee, Ko, and Shin}{Lee et~al\mbox{.}}{2020}]%
        {lee2020hypergraph}
\bibfield{author}{\bibinfo{person}{Geon Lee}, \bibinfo{person}{Jihoon Ko},
  {and} \bibinfo{person}{Kijung Shin}.} \bibinfo{year}{2020}\natexlab{}.
\newblock \showarticletitle{Hypergraph motifs: concepts, algorithms, and
  discoveries}.
\newblock \bibinfo{journal}{\emph{PVLDB}} \bibinfo{volume}{13},
  \bibinfo{number}{12} (\bibinfo{year}{2020}), \bibinfo{pages}{2256--2269}.
\newblock


\bibitem[\protect\citeauthoryear{Lee, Lee, Kim, Kosiorek, Choi, and Teh}{Lee
  et~al\mbox{.}}{2019}]%
        {lee2019set}
\bibfield{author}{\bibinfo{person}{Juho Lee}, \bibinfo{person}{Yoonho Lee},
  \bibinfo{person}{Jungtaek Kim}, \bibinfo{person}{Adam Kosiorek},
  \bibinfo{person}{Seungjin Choi}, {and} \bibinfo{person}{Yee~Whye Teh}.}
  \bibinfo{year}{2019}\natexlab{}.
\newblock \showarticletitle{Set transformer: A framework for attention-based
  permutation-invariant neural networks}. In \bibinfo{booktitle}{\emph{ICML}}.
\newblock


\bibitem[\protect\citeauthoryear{Li, He, and Zhu}{Li et~al\mbox{.}}{2018}]%
        {li2018tail}
\bibfield{author}{\bibinfo{person}{Jianbo Li}, \bibinfo{person}{Jingrui He},
  {and} \bibinfo{person}{Yada Zhu}.} \bibinfo{year}{2018}\natexlab{}.
\newblock \showarticletitle{E-tail product return prediction via
  hypergraph-based local graph cut}. In \bibinfo{booktitle}{\emph{KDD}}.
\newblock


\bibitem[\protect\citeauthoryear{Li, Wang, Wang, and Leskovec}{Li
  et~al\mbox{.}}{2020}]%
        {li2020distance}
\bibfield{author}{\bibinfo{person}{Pan Li}, \bibinfo{person}{Yanbang Wang},
  \bibinfo{person}{Hongwei Wang}, {and} \bibinfo{person}{Jure Leskovec}.}
  \bibinfo{year}{2020}\natexlab{}.
\newblock \showarticletitle{Distance encoding: Design provably more powerful
  neural networks for graph representation learning}. In
  \bibinfo{booktitle}{\emph{NeurIPS}}.
\newblock


\bibitem[\protect\citeauthoryear{Lin}{Lin}{1991}]%
        {lin1991divergence}
\bibfield{author}{\bibinfo{person}{Jianhua Lin}.}
  \bibinfo{year}{1991}\natexlab{}.
\newblock \showarticletitle{Divergence measures based on the Shannon entropy}.
\newblock \bibinfo{journal}{\emph{IEEE Transactions on Information theory}}
  \bibinfo{volume}{37}, \bibinfo{number}{1} (\bibinfo{year}{1991}),
  \bibinfo{pages}{145--151}.
\newblock


\bibitem[\protect\citeauthoryear{Ma, Rabbany, and Romero-Soriano}{Ma
  et~al\mbox{.}}{2021}]%
        {ma2021graph}
\bibfield{author}{\bibinfo{person}{Liheng Ma}, \bibinfo{person}{Reihaneh
  Rabbany}, {and} \bibinfo{person}{Adriana Romero-Soriano}.}
  \bibinfo{year}{2021}\natexlab{}.
\newblock \showarticletitle{Graph attention networks with positional
  embeddings}. In \bibinfo{booktitle}{\emph{PAKDD}}.
\newblock


\bibitem[\protect\citeauthoryear{Mialon, Chen, Selosse, and Mairal}{Mialon
  et~al\mbox{.}}{2021}]%
        {mialon2021graphit}
\bibfield{author}{\bibinfo{person}{Gr{\'e}goire Mialon},
  \bibinfo{person}{Dexiong Chen}, \bibinfo{person}{Margot Selosse}, {and}
  \bibinfo{person}{Julien Mairal}.} \bibinfo{year}{2021}\natexlab{}.
\newblock \showarticletitle{Graphit: Encoding graph structure in transformers}.
\newblock \bibinfo{journal}{\emph{arXiv:2106.05667}} (\bibinfo{year}{2021}).
\newblock


\bibitem[\protect\citeauthoryear{Mikolov, Chen, Corrado, and Dean}{Mikolov
  et~al\mbox{.}}{2013a}]%
        {mikolov2013efficient}
\bibfield{author}{\bibinfo{person}{Tomas Mikolov}, \bibinfo{person}{Kai Chen},
  \bibinfo{person}{Greg Corrado}, {and} \bibinfo{person}{Jeffrey Dean}.}
  \bibinfo{year}{2013}\natexlab{a}.
\newblock \showarticletitle{Efficient estimation of word representations in
  vector space}.
\newblock \bibinfo{journal}{\emph{arXiv:1301.3781}} (\bibinfo{year}{2013}).
\newblock


\bibitem[\protect\citeauthoryear{Mikolov, Sutskever, Chen, Corrado, and
  Dean}{Mikolov et~al\mbox{.}}{2013b}]%
        {mikolov2013distributed}
\bibfield{author}{\bibinfo{person}{Tomas Mikolov}, \bibinfo{person}{Ilya
  Sutskever}, \bibinfo{person}{Kai Chen}, \bibinfo{person}{Greg~S Corrado},
  {and} \bibinfo{person}{Jeff Dean}.} \bibinfo{year}{2013}\natexlab{b}.
\newblock \showarticletitle{Distributed representations of words and phrases
  and their compositionality}. In \bibinfo{booktitle}{\emph{NeurIPS}}.
\newblock


\bibitem[\protect\citeauthoryear{Molloy}{Molloy}{2005}]%
        {molloy2005cores}
\bibfield{author}{\bibinfo{person}{Michael Molloy}.}
  \bibinfo{year}{2005}\natexlab{}.
\newblock \showarticletitle{Cores in random hypergraphs and Boolean formulas}.
\newblock \bibinfo{journal}{\emph{Random Structures \& Algorithms}}
  \bibinfo{volume}{27}, \bibinfo{number}{1} (\bibinfo{year}{2005}),
  \bibinfo{pages}{124--135}.
\newblock


\bibitem[\protect\citeauthoryear{Niepert, Ahmed, and Kutzkov}{Niepert
  et~al\mbox{.}}{2016}]%
        {niepert2016learning}
\bibfield{author}{\bibinfo{person}{Mathias Niepert}, \bibinfo{person}{Mohamed
  Ahmed}, {and} \bibinfo{person}{Konstantin Kutzkov}.}
  \bibinfo{year}{2016}\natexlab{}.
\newblock \showarticletitle{Learning convolutional neural networks for graphs}.
  In \bibinfo{booktitle}{\emph{ICML}}.
\newblock


\bibitem[\protect\citeauthoryear{Page, Brin, Motwani, and Winograd}{Page
  et~al\mbox{.}}{1999}]%
        {page1999pagerank}
\bibfield{author}{\bibinfo{person}{Lawrence Page}, \bibinfo{person}{Sergey
  Brin}, \bibinfo{person}{Rajeev Motwani}, {and} \bibinfo{person}{Terry
  Winograd}.} \bibinfo{year}{1999}\natexlab{}.
\newblock \bibinfo{booktitle}{\emph{The PageRank citation ranking: Bringing
  order to the web.}}
\newblock \bibinfo{type}{{T}echnical {R}eport}. \bibinfo{institution}{Stanford
  InfoLab}.
\newblock


\bibitem[\protect\citeauthoryear{Paranjape, Benson, and Leskovec}{Paranjape
  et~al\mbox{.}}{2017}]%
        {paranjape2017motifs}
\bibfield{author}{\bibinfo{person}{Ashwin Paranjape}, \bibinfo{person}{Austin~R
  Benson}, {and} \bibinfo{person}{Jure Leskovec}.}
  \bibinfo{year}{2017}\natexlab{}.
\newblock \showarticletitle{Motifs in temporal networks}. In
  \bibinfo{booktitle}{\emph{WSDM}}. \bibinfo{pages}{601--610}.
\newblock


\bibitem[\protect\citeauthoryear{Park, Chang, Lee, Kim, et~al\mbox{.}}{Park
  et~al\mbox{.}}{2022}]%
        {park2022grpe}
\bibfield{author}{\bibinfo{person}{Wonpyo Park}, \bibinfo{person}{Woong-Gi
  Chang}, \bibinfo{person}{Donggeon Lee}, \bibinfo{person}{Juntae Kim},
  {et~al\mbox{.}}} \bibinfo{year}{2022}\natexlab{}.
\newblock \showarticletitle{GRPE: Relative Positional Encoding for Graph
  Transformer}. In \bibinfo{booktitle}{\emph{ICLR Machine Learning for Drug
  Discovery Workshop}}.
\newblock


\bibitem[\protect\citeauthoryear{Qiu, Huang, Chen, and Yin}{Qiu
  et~al\mbox{.}}{2021}]%
        {qiu2021exploiting}
\bibfield{author}{\bibinfo{person}{Ruihong Qiu}, \bibinfo{person}{Zi Huang},
  \bibinfo{person}{Tong Chen}, {and} \bibinfo{person}{Hongzhi Yin}.}
  \bibinfo{year}{2021}\natexlab{}.
\newblock \showarticletitle{Exploiting positional information for session-based
  recommendation}.
\newblock \bibinfo{journal}{\emph{TOIS}} \bibinfo{volume}{40},
  \bibinfo{number}{2} (\bibinfo{year}{2021}), \bibinfo{pages}{1--24}.
\newblock


\bibitem[\protect\citeauthoryear{Raffel, Shazeer, Roberts, Lee, Narang, Matena,
  Zhou, Li, Liu, et~al\mbox{.}}{Raffel et~al\mbox{.}}{2020}]%
        {raffel2020exploring}
\bibfield{author}{\bibinfo{person}{Colin Raffel}, \bibinfo{person}{Noam
  Shazeer}, \bibinfo{person}{Adam Roberts}, \bibinfo{person}{Katherine Lee},
  \bibinfo{person}{Sharan Narang}, \bibinfo{person}{Michael Matena},
  \bibinfo{person}{Yanqi Zhou}, \bibinfo{person}{Wei Li},
  \bibinfo{person}{Peter~J Liu}, {et~al\mbox{.}}}
  \bibinfo{year}{2020}\natexlab{}.
\newblock \showarticletitle{Exploring the limits of transfer learning with a
  unified text-to-text transformer.}
\newblock \bibinfo{journal}{\emph{JMLR}} \bibinfo{volume}{21},
  \bibinfo{number}{140} (\bibinfo{year}{2020}), \bibinfo{pages}{1--67}.
\newblock


\bibitem[\protect\citeauthoryear{Sato, Yamada, and Kashima}{Sato
  et~al\mbox{.}}{2021}]%
        {sato2021random}
\bibfield{author}{\bibinfo{person}{Ryoma Sato}, \bibinfo{person}{Makoto
  Yamada}, {and} \bibinfo{person}{Hisashi Kashima}.}
  \bibinfo{year}{2021}\natexlab{}.
\newblock \showarticletitle{Random features strengthen graph neural networks}.
  In \bibinfo{booktitle}{\emph{SDM}}.
\newblock


\bibitem[\protect\citeauthoryear{Shaw, Uszkoreit, and Vaswani}{Shaw
  et~al\mbox{.}}{2018}]%
        {shaw2018self}
\bibfield{author}{\bibinfo{person}{Peter Shaw}, \bibinfo{person}{Jakob
  Uszkoreit}, {and} \bibinfo{person}{Ashish Vaswani}.}
  \bibinfo{year}{2018}\natexlab{}.
\newblock \showarticletitle{Self-attention with relative position
  representations}. In \bibinfo{booktitle}{\emph{NAACL}}.
\newblock


\bibitem[\protect\citeauthoryear{Swati, Ashish, Nitish, Rohan, and
  Denzil}{Swati et~al\mbox{.}}{2016}]%
        {Swati2016dblp}
\bibfield{author}{\bibinfo{person}{A Swati}, \bibinfo{person}{S Ashish},
  \bibinfo{person}{M Nitish}, \bibinfo{person}{K Rohan}, {and}
  \bibinfo{person}{C Denzil}.} \bibinfo{year}{2016}\natexlab{}.
\newblock \bibinfo{title}{Dblp records and entries for key computer science
  conferences}.
\newblock
\newblock


\bibitem[\protect\citeauthoryear{Tudisco and Higham}{Tudisco and
  Higham}{2021}]%
        {tudisco2021node}
\bibfield{author}{\bibinfo{person}{Francesco Tudisco} {and}
  \bibinfo{person}{Desmond~J Higham}.} \bibinfo{year}{2021}\natexlab{}.
\newblock \showarticletitle{Node and edge nonlinear eigenvector centrality for
  hypergraphs}.
\newblock \bibinfo{journal}{\emph{Communications Physics}} \bibinfo{volume}{4},
  \bibinfo{number}{1} (\bibinfo{year}{2021}), \bibinfo{pages}{201}.
\newblock


\bibitem[\protect\citeauthoryear{Tudisco and Higham}{Tudisco and
  Higham}{2023}]%
        {tudisco2023core}
\bibfield{author}{\bibinfo{person}{Francesco Tudisco} {and}
  \bibinfo{person}{Desmond~J Higham}.} \bibinfo{year}{2023}\natexlab{}.
\newblock \showarticletitle{Core-periphery detection in hypergraphs}.
\newblock \bibinfo{journal}{\emph{SIMODS}} \bibinfo{volume}{5},
  \bibinfo{number}{1} (\bibinfo{year}{2023}), \bibinfo{pages}{1--21}.
\newblock


\bibitem[\protect\citeauthoryear{Vaswani, Shazeer, Parmar, Uszkoreit, Jones,
  Gomez, Kaiser, and Polosukhin}{Vaswani et~al\mbox{.}}{2017}]%
        {vaswani2017attention}
\bibfield{author}{\bibinfo{person}{Ashish Vaswani}, \bibinfo{person}{Noam
  Shazeer}, \bibinfo{person}{Niki Parmar}, \bibinfo{person}{Jakob Uszkoreit},
  \bibinfo{person}{Llion Jones}, \bibinfo{person}{Aidan~N Gomez},
  \bibinfo{person}{{\L}ukasz Kaiser}, {and} \bibinfo{person}{Illia
  Polosukhin}.} \bibinfo{year}{2017}\natexlab{}.
\newblock \showarticletitle{Attention is all you need}. In
  \bibinfo{booktitle}{\emph{NeurIPS}}.
\newblock


\bibitem[\protect\citeauthoryear{Wang, Yin, Zhang, and Li}{Wang
  et~al\mbox{.}}{2022}]%
        {wang2022equivariant}
\bibfield{author}{\bibinfo{person}{Haorui Wang}, \bibinfo{person}{Haoteng Yin},
  \bibinfo{person}{Muhan Zhang}, {and} \bibinfo{person}{Pan Li}.}
  \bibinfo{year}{2022}\natexlab{}.
\newblock \showarticletitle{Equivariant and stable positional encoding for more
  powerful graph neural networks}. In \bibinfo{booktitle}{\emph{ICLR}}.
\newblock


\bibitem[\protect\citeauthoryear{Wang, Zheng, Ye, Gan, Li, Song, Zhou, Ma, Yu,
  Gai, et~al\mbox{.}}{Wang et~al\mbox{.}}{2019}]%
        {wang2019deep}
\bibfield{author}{\bibinfo{person}{Minjie Wang}, \bibinfo{person}{Da Zheng},
  \bibinfo{person}{Zihao Ye}, \bibinfo{person}{Quan Gan},
  \bibinfo{person}{Mufei Li}, \bibinfo{person}{Xiang Song},
  \bibinfo{person}{Jinjing Zhou}, \bibinfo{person}{Chao Ma},
  \bibinfo{person}{Lingfan Yu}, \bibinfo{person}{Yu Gai}, {et~al\mbox{.}}}
  \bibinfo{year}{2019}\natexlab{}.
\newblock \showarticletitle{Deep graph library: A graph-centric,
  highly-performant package for graph neural networks}.
\newblock \bibinfo{journal}{\emph{arXiv:1909.01315}} (\bibinfo{year}{2019}).
\newblock


\bibitem[\protect\citeauthoryear{Wolf, Klinvex, and Dunlavy}{Wolf
  et~al\mbox{.}}{2016}]%
        {wolf2016advantages}
\bibfield{author}{\bibinfo{person}{Michael~M Wolf}, \bibinfo{person}{Alicia~M
  Klinvex}, {and} \bibinfo{person}{Daniel~M Dunlavy}.}
  \bibinfo{year}{2016}\natexlab{}.
\newblock \showarticletitle{Advantages to modeling relational data using
  hypergraphs versus graphs}. In \bibinfo{booktitle}{\emph{HPEC}}.
\newblock


\bibitem[\protect\citeauthoryear{Xia, Yin, Yu, Wang, Cui, and Zhang}{Xia
  et~al\mbox{.}}{2021}]%
        {xia2021self}
\bibfield{author}{\bibinfo{person}{Xin Xia}, \bibinfo{person}{Hongzhi Yin},
  \bibinfo{person}{Junliang Yu}, \bibinfo{person}{Qinyong Wang},
  \bibinfo{person}{Lizhen Cui}, {and} \bibinfo{person}{Xiangliang Zhang}.}
  \bibinfo{year}{2021}\natexlab{}.
\newblock \showarticletitle{Self-supervised hypergraph convolutional networks
  for session-based recommendation}. In \bibinfo{booktitle}{\emph{AAAI}}.
\newblock


\bibitem[\protect\citeauthoryear{Yadati, Nimishakavi, Yadav, Nitin, Louis, and
  Talukdar}{Yadati et~al\mbox{.}}{2019}]%
        {yadati2019hypergcn}
\bibfield{author}{\bibinfo{person}{Naganand Yadati}, \bibinfo{person}{Madhav
  Nimishakavi}, \bibinfo{person}{Prateek Yadav}, \bibinfo{person}{Vikram
  Nitin}, \bibinfo{person}{Anand Louis}, {and} \bibinfo{person}{Partha
  Talukdar}.} \bibinfo{year}{2019}\natexlab{}.
\newblock \showarticletitle{Hypergcn: A new method for training graph
  convolutional networks on hypergraphs}. In
  \bibinfo{booktitle}{\emph{NeurIPS}}.
\newblock


\bibitem[\protect\citeauthoryear{Ying, Cai, Luo, Zheng, Ke, He, Shen, and
  Liu}{Ying et~al\mbox{.}}{2021}]%
        {ying2021transformers}
\bibfield{author}{\bibinfo{person}{Chengxuan Ying}, \bibinfo{person}{Tianle
  Cai}, \bibinfo{person}{Shengjie Luo}, \bibinfo{person}{Shuxin Zheng},
  \bibinfo{person}{Guolin Ke}, \bibinfo{person}{Di He},
  \bibinfo{person}{Yanming Shen}, {and} \bibinfo{person}{Tie-Yan Liu}.}
  \bibinfo{year}{2021}\natexlab{}.
\newblock \showarticletitle{Do transformers really perform badly for graph
  representation?}. In \bibinfo{booktitle}{\emph{NeurIPS}}.
\newblock


\bibitem[\protect\citeauthoryear{Yoon, Song, Shin, and Yi}{Yoon
  et~al\mbox{.}}{2020}]%
        {yoon2020much}
\bibfield{author}{\bibinfo{person}{Se-eun Yoon}, \bibinfo{person}{Hyungseok
  Song}, \bibinfo{person}{Kijung Shin}, {and} \bibinfo{person}{Yung Yi}.}
  \bibinfo{year}{2020}\natexlab{}.
\newblock \showarticletitle{How much and when do we need higher-order
  information in hypergraphs? a case study on hyperedge prediction}. In
  \bibinfo{booktitle}{\emph{WWW}}.
\newblock


\bibitem[\protect\citeauthoryear{You, Ying, and Leskovec}{You
  et~al\mbox{.}}{2019}]%
        {you2019position}
\bibfield{author}{\bibinfo{person}{Jiaxuan You}, \bibinfo{person}{Rex Ying},
  {and} \bibinfo{person}{Jure Leskovec}.} \bibinfo{year}{2019}\natexlab{}.
\newblock \showarticletitle{Position-aware graph neural networks}. In
  \bibinfo{booktitle}{\emph{ICML}}.
\newblock


\bibitem[\protect\citeauthoryear{Zhang, Zhang, Xia, and Sun}{Zhang
  et~al\mbox{.}}{2020a}]%
        {zhang2020graph}
\bibfield{author}{\bibinfo{person}{Jiawei Zhang}, \bibinfo{person}{Haopeng
  Zhang}, \bibinfo{person}{Congying Xia}, {and} \bibinfo{person}{Li Sun}.}
  \bibinfo{year}{2020}\natexlab{a}.
\newblock \showarticletitle{Graph-bert: Only attention is needed for learning
  graph representations}.
\newblock \bibinfo{journal}{\emph{arXiv:2001.05140}} (\bibinfo{year}{2020}).
\newblock


\bibitem[\protect\citeauthoryear{Zhang, Zou, and Ma}{Zhang
  et~al\mbox{.}}{2020b}]%
        {zhang2019hyper}
\bibfield{author}{\bibinfo{person}{R Zhang}, \bibinfo{person}{Y Zou}, {and}
  \bibinfo{person}{J Ma}.} \bibinfo{year}{2020}\natexlab{b}.
\newblock \showarticletitle{Hyper-SAGNN: a self-attention based graph neural
  network for hypergraphs}. In \bibinfo{booktitle}{\emph{ICLR}}.
\newblock


\bibitem[\protect\citeauthoryear{Zhao, Li, Wen, Wang, Liu, Sun, Xie, and
  Ye}{Zhao et~al\mbox{.}}{2021}]%
        {zhao2021gophormer}
\bibfield{author}{\bibinfo{person}{Jianan Zhao}, \bibinfo{person}{Chaozhuo Li},
  \bibinfo{person}{Qianlong Wen}, \bibinfo{person}{Yiqi Wang},
  \bibinfo{person}{Yuming Liu}, \bibinfo{person}{Hao Sun},
  \bibinfo{person}{Xing Xie}, {and} \bibinfo{person}{Yanfang Ye}.}
  \bibinfo{year}{2021}\natexlab{}.
\newblock \showarticletitle{Gophormer: Ego-Graph Transformer for Node
  Classification}.
\newblock \bibinfo{journal}{\emph{arXiv:2110.13094}} (\bibinfo{year}{2021}).
\newblock


\end{thebibliography}

    \appendix
    \begin{table*}[t]
    \vspace{-2mm}
    \begin{center}
        \caption{\label{table:searchspace} Search spaces of hyperparameters.}
        \scalebox{0.78}{
            \begin{tabular}{c |c c c c|c c c c c c}
            \toprule
            \multirow{2}{*}{Hyperparameter} && \multicolumn{2}{c}{Datasets for Problem~\ref{problem}} &&& \multicolumn{5}{c}{Datasets for Downstream Tasks} \\
            \cline{3-4} \cline{7-11} 
            && Coauth-AMiner & The others &&& Ranking Halo & Ranking H-Index & Clustering DBLP & Clustering AMiner & Synthetic E-tail \\ 
            \midrule
            Learning rate && 0.001, 0.0001 & 0.001, 0.0001 &&& 0.0001, 0.001 & 0.0001, 0.0005 & 0.0001, 0.0005, 0.005 & 0.0001, 0.0005 & 0.001, 0.003, 0.005, 0.01, 0.03, 0.05 \\
            Size of batch && 256, 512 & 64, 128 &&& 128, 256 & 32, 64, 128 & 32, 64, 128 & 32, 64, 128 & 64, 128, 256, 512 \\
            Number of layers && $1, 2^{*}$ & 1, 2 &&& 1 & 1, 2 & 1, 2 & 1, 2 & 1, 2 \\
            \bottomrule
        \end{tabular}}
    \end{center}
    \small
    *: We use $1$ layer for \method, AST, and HST.
\end{table*}
\section{Parameter Setting}\label{appendix:hp}
The search spaces of hyperparameters are given in Table~\ref{table:searchspace}.
For all models, we fix the hidden dimension to $64$, the final embedding dimension to $128$, the number of the inducing points to $4$, the number of attention layers to $2$, and the dropout ratio to $0.7$.

We employ sampling for efficiency in hyperedge-to-node message passing (but not in node-to-hyperedge message passing) and set the size of sampling (i.e. a set of hyperedges sampled among those incidents to each node) as follows: (a) All methods do not use sampling in Coauthorship-DBLP, (b) HST, AST, and \method sample $40$ hyperedges in the other datasets, and (c) HNHN, HGNN, HCHA, HAT, and UniGCNII do not sample in StackOverflow-Biology but sample $40$ or $100$ hyperedges in the other datasets.

Exceptionally, for HCHA~\cite{bai2021hypergraph} and HNN~\cite{aponte2022hypergraph}, we use full-batch training without sampling. For a fair comparison, we tune their hyperparameters in a larger search space: $\{0.005, 0,01, 0.03, 0.05, 0.1\}$ for learning rates, $\{1,2\}$ for the number of layers, and $\{0.3, 0.5, 0.7\}$ for dropout ratios, while we fix the dimension of final node and hyperedge embeddings to $128$, the hidden dimension to $64$, and the number of epochs to $300$ with early stopping.

\begin{table*}[t]
    \vspace{-2mm}
    \begin{center}
        \caption{\label{tab:ablation:pe} Comparison of positional encoding schemes. The symbol $\mathbf{I}_w$ denotes inducing points.}
        \scalebox{0.78}{
            \begin{tabular}{c c | c c c c c c c c||c c c}
            \toprule
            \multicolumn{2}{c|}{\multirow{2}{*}{Dataset}} & & \multicolumn{6}{c}{Positional Encodings} &&& \multicolumn{2}{c}{\PE} \\
            \cline{4-9} \cline{12-13}
            && w/o PE & GraphIT/DK & GraphIT/PRWK & Shaw/DK & Shaw/PRWK & LSPE & WholeOrderPE &&& w/o $\mathbf{I}_w$ & w/ $\mathbf{I}_w$  \\ 
            \midrule
            \multirow{2}{*}{\shortstack{Stack\\Biology}} & MicroF1 & 0.732 $\pm$ 0.002 & 0.719 $\pm$ 0.006 & 0.710 $\pm$ 0.004 & 0.731 $\pm$ 0.003 & 0.456 $\pm$ 0.014  & 0.727 $\pm$ 0.003 & 0.732 $\pm$ 0.003 &&& \textbf{0.737} $\pm$ 0.006 & \textbf{0.737} $\pm$ 0.003 \\
            & MacroF1 & 0.672 $\pm$ 0.004 & 0.645 $\pm$ 0.012 & 0.638 $\pm$ 0.012 & 0.667 $\pm$ 0.005 & 0.338 $\pm$ 0.038 & 0.658 $\pm$ 0.010 & 0.669 $\pm$ 0.011 &&& \textbf{0.680} $\pm$ 0.005 & 0.679 $\pm$ 0.007 \\
            \bottomrule
        \end{tabular}}
    \end{center}
	\vspace{-2mm}
\end{table*}

\begin{table*}[t]
    \vspace{-2mm}
    \begin{center}
        \caption{\label{tab:ablation:cent} Comparison of node-centrality measures that can be used for \PE.}
        \scalebox{0.785}{
            \begin{tabular}{c c | c c c c |c c c c |c c c}
            \toprule
            \multicolumn{2}{c|}{Dataset} & Degree & Coreness & Eigenvector & PageRank && H-Coreness & H-Eigenvector & Vector Centrality && \method & \method-all  \\ 
            \midrule
            \multirow{2}{*}{\shortstack{Stack\\Biology}} & MicroF1 & 0.735 $\pm$ 0.001 & 0.739 $\pm$ 0.005 & 0.738 $\pm$ 0.003 & 0.741 $\pm$ 0.003 
            && 0.738 $\pm$ 0.005 & 0.738 $\pm$ 0.004 & \textbf{0.745} $\pm$ 0.002 
            && \textit{0.742} $\pm$ 0.003 & \textbf{0.745} $\pm$ 0.002\\
            & MacroF1 & 0.678 $\pm$ 0.004 & 0.685 $\pm$ 0.010 & 0.681 $\pm$ 0.004 & 0.681 $\pm$ 0.004 
            && 0.683 $\pm$ 0.008 & 0.681 $\pm$ 0.006 & \textit{0.687} $\pm$ 0.003 
            && 0.686 $\pm$ 0.004 & \textbf{0.690} $\pm$ 0.005\\ 
            \bottomrule
        \end{tabular}}
    \end{center}
    \vspace{-2mm}
\end{table*}


\begin{table*}[t]
    \vspace{-2mm}
    \begin{center}
        \caption{\label{tab:ablation:ind_arch} Effect of the number of inducing points and comparison of architectures for hyperedge-to-node message passing.} 
        \scalebox{0.8}{
            \begin{tabular}{c c |c c c c c|| c c c c c}
            \toprule
            \multicolumn{2}{c|}{\multirow{2}{*}{Dataset}} && \multicolumn{3}{c}{Number of Inducing Points} && & & \multicolumn{2}{c}{MSG. Passing from Hyperedge to Node} &  \\ 
            \cline{4-6} \cline{10-11}
            & && 2 & 4 & 8 && HNHN & HAT & \method + HNHN & \method + HAT & \method \\ 
            \midrule
            \multirow{2}{*}{\shortstack{Stack\\Biology}} & MicroF1 &&
            0.740 $\pm$ 0.002 & \textbf{0.742} $\pm$ 0.003 & \textbf{0.742} $\pm$ 0.003 &&
            0.640 $\pm$ 0.005 & 0.661 $\pm$ 0.005 & 0.645 $\pm$ 0.023 & 0.670 $\pm$ 0.015  & \textbf{0.742} $\pm$ 0.003\\
            & MacroF1 && 
            0.682 $\pm$ 0.006 & 0.686 $\pm$ 0.004 & \textbf{0.688} $\pm$ 0.002 &&
            0.592 $\pm$ 0.006 & 0.606 $\pm$ 0.005 & 0.583 $\pm$ 0.029 & 0.618 $\pm$ 0.010  & \textbf{0.686} $\pm$ 0.004\\ 
            \bottomrule
        \end{tabular}}
    \end{center}
\end{table*}

\section{Additional Ablation Studies}\label{appendix:additional_abl}

\subsection{Positional Encoding}\label{appendix:abl_pe}
We examine the usefulness of \PE by replacing it with alternative positional encoding schemes for graph neural networks (see Section~\ref{sec:pe}).
In our study, we initialize node features using random walks (see Section ~\ref{sec:exp:setting}) so that they capture global positional information. Thus,
we compare \PE with two relative positional encodings: Shaw et al.~\cite{shaw2018self} and GraphIT~\cite{mialon2021graphit}, using diffusion kernels~\cite{kondor2004diffusion} (DK) and the p-step random-walk kernel~\cite{mialon2021graphit} (PRWK) for calculating distances between nodes.
We also consider learnable positional encodings called LSPE~\cite{dwivedi2021graph}, which are added to node embeddings (as \PE is added) and updated together with other parameters.
In addition, we consider \textit{WholeOrderPE}, a variant of \PE that uses the global centrality order among all nodes, instead of the relative order within each hyperedge.
Since the relative positional encoding methods pose challenges when used with inducing points, we do not employ inducing points consistently, and for \method, we consider two versions with and without inducing points.
The base model for comparing positional encoding schemes is defined as follows:
\begin{equation}\label{equation:abl_pe}
    \tilde{V}_{e}^{(l)} = MAB(V_{e}^{(l-1)}, V_{e}^{(l-1)} ; PE)
\end{equation}
\begin{equation}
    H_e^{(l)} = \MultiheadAtt(H_e^{(l-1)}, \tilde{V}_{e}^{(l)})
\end{equation}
\begin{equation}\label{eqution:abl_pe_e}
    X_v^{(l)} = \MultiheadAtt(X_v^{(l-1)}, \WithinAttention(E_{v}^{(l)}))
\end{equation}
Only Eq.~\eqref{equation:abl_pe} depends on positional encoding schemes $PE$. 
To specifically investigate the impact of node positional encoding schemes, we do not employ any positional encodings for hyperedges.


The results are shown in Table~\ref{tab:ablation:pe}, where \PE outperforms all other positional encoding methods, particularly surpassing WholeOrderPE. This highlights the effectiveness of incorporating relative order within hyperedges for positional encodings. It is worth noting that \method with inducing points, which offers improved efficiency, achieves competitive performance compared to \method without inducing points.

\subsection{Node-centrality measures} \label{appendix:abl_cent}
We investigate the impact of node-centrality measures used in \PE on the performance of \method. We consider individually the following seven node-centrality measures:
\begin{itemize}[leftmargin=*]
    
    \item \textbf{Degree}: The number of hyperedges a node belongs to.
    
    \item \textbf{Coreness}~\cite{molloy2005cores, jiang2017parallel, bu2023hypercore}: 
    The maximum $k$ such that a node belongs to the $k$-core which is the maximal sub-hypergraph where every node has at least degree $k$ within it.

    \item \textbf{Eigenvector Centrality}~\cite{bonacich1987power} and \textbf{PageRank}~\cite{page1999pagerank}: The eigenvector centrality and the PageRank score on a clique-expanded graph where each edge weight $w_{u, v} = |\{e \in \SE : v \in e \text{ and } u \in e \}|$.


    \item \textbf{H-Coreness}~\cite{tudisco2023core}: Hypergraph core-periphery scores determining the proximity of nodes to the hypergraph core.

    \item \textbf{H-Eigenvector}~\cite{tudisco2021node}: Node and hyperedge centralities on hypergraphs using the max centrality model. Different positional encodings are used for different message-passing directions.

    \item \textbf{Vector Centrality}~\cite{kovalenko2022vector}: A vectorial measure of the roles of each node at different orders of interactions. The dimension of this measure is one less than the maximum hyperedge size.
    
\end{itemize}
In \method, we use only four centrality measures (spec., degree, coreness, eigenvector, and PageRank) for computational efficiency (see Appendix~\ref{appendix:complexity} for details).
We additionally consider \method-all, which uses all seven centrality measures together.
Table~\ref{tab:ablation:cent} shows that the performance of \method is not highly sensitive to the choice of node centrality measures, and especially it consistently outperforms HST (the strongest baseline approach) regardless of the choice of centrality measures.
Moreover, the results indicate that increasing the dimensionality of \PE tends to lead to performance improvement, which is evident from the high performance of vector centrality, \method, and \method-all, which utilize higher-dimensional \PE.


\subsection{Advantage of using \WithinAttention for both directions}\label{appendix:abl_partial}
Recall that we devise the \WithinAttention with the intention of capturing edge-dependent relationships between nodes within each hyperedge. However, we also utilize this module for the propagation from hyperedges to nodes.  
To validate the effectiveness of using \WithinAttention in both directions,
we replace the message passing from hyperedges to nodes with the aggregation methods used in HNHN (\method + HNHN) and HAT (\method + HAT).
Table~\ref{tab:ablation:ind_arch} demonstrates that \method outperforms both \method + HNHN and \method + HAT, providing justification for employing the \WithinAttention module in both directions. It is also worth noting that \method + HNHN and \method + HAT outperform their corresponding vanilla models, HNHN and HAT.


\begin{table}[t]
    \begin{center}
        \caption{Effect of inputs to the final classifier\label{tab:classifier}}
        \scalebox{0.83}{
            \begin{tabular}{l l| c c c}
            \toprule
            Dataset & Metric & Best Competitor &  \method-IM & \method\\
            \midrule
            \multirow{2}{*}{\shortstack[l]{Coauth-\\DBLP}} & MicroF1 & 0.564 $\pm$ 0.004 & 0.602 $\pm$ 0.002 & \textbf{0.605} $\pm$ 0.002 \\
             & MacroF1 & 0.549 $\pm$ 0.003 & 0.592 $\pm$ 0.002 & \textbf{0.595} $\pm$ 0.002 \\
            \midrule
            \multirow{2}{*}{\shortstack[l]{Coauth-\\AMiner}} & MicroF1 & 0.596 $\pm$ 0.007 & \textbf{0.637} $\pm$ 0.003 & 0.630 $\pm$ 0.005 \\
             & MacroF1 & 0.583 $\pm$ 0.008 & \textbf{0.631} $\pm$ 0.003 & 0.623 $\pm$ 0.007 \\
            \midrule
            \multirow{2}{*}{\shortstack[l]{Email-\\Enron}} & MicroF1 & 0.779 $\pm$ 0.067 & \textbf{0.858} $\pm$ 0.001 & 0.826 $\pm$ 0.001 \\
             & MacroF1 & 0.681 $\pm$ 0.123 & \textbf{0.796} $\pm$ 0.004 & 0.760 $\pm$ 0.004 \\
             \midrule
            \multirow{2}{*}{\shortstack[l]{Email-\\Eu}} & MicroF1 & 0.671 $\pm$ 0.001 & \textbf{0.687} $\pm$ 0.004 & 0.671 $\pm$ 0.000 \\
             & MacroF1 & 0.640 $\pm$ 0.002 & \textbf{0.660} $\pm$ 0.009 & 0.646 $\pm$ 0.003 \\
            \midrule
            \multirow{2}{*}{\shortstack[l]{Stack-\\Biology}} & MicroF1 & 0.694 $\pm$ 0.002 & 0.736 $\pm$ 0.003 & \textbf{0.742} $\pm$ 0.003 \\
             & MacroF1 & 0.631 $\pm$ 0.006 & 0.679 $\pm$ 0.007 & \textbf{0.686} $\pm$ 0.004 \\
            \midrule
            \multirow{2}{*}{\shortstack[l]{Stack-\\Physics}} & MicroF1 & 0.755 $\pm$ 0.010 & 0.769 $\pm$ 0.001 & \textbf{0.770} $\pm$ 0.003 \\
             & MacroF1 & 0.666 $\pm$ 0.013 & 0.692 $\pm$ 0.011 & \textbf{0.707} $\pm$ 0.004 \\
            \bottomrule
        \end{tabular}}
    \end{center}
\end{table}

\subsection{The number of inducing points}\label{appendix:abl_ind_cnet}
We explore how the performance of \method changes with respect to the numbers of inducing points (2, 4, and 8). 
Table~\ref{tab:ablation:ind_arch} shows that the performance improves as more inducing points are employed. However, considering memory constraints, using 4 inducing points is favorable as it requires half the memory while achieving comparable performance to using 8 inducing points.

\subsection{Inputs to the final classifier}
As discussed in Section~\ref{method:classifier}, we consider two possible inputs for a single-layer perceptron classifier: 
(a) intermediate edge-dependent node embeddings, referred to as \method-IM, which are generated by \WithinAttention inside a \method (Eq.~\eqref{eq:classifier2}) and (b) the concatenation of final output node and hyperedge embeddings from \method (Eq.~\eqref{eq:classifier}).
Table~\ref{tab:classifier} indicates that one approach does not consistently outperform the other; \method-IM performs better in Coauth-AMiner, Email-Enron, and Email-Eu datasets, while \method performs better in the remaining three datasets.
It is also worth noting that both approaches are superior to HST, which is overall the best competitor.


\section{Time complexity of node centrality measures }\label{appendix:complexity}

We analyze the time complexity of computing each node's centrality using each of the four node measures used for \PE in \method.
We assume $O(|\SV|+|\SE|)\in O(\sum_{e \in \SE} |e|)$ for a hypergraph $\SG = (\SV,\SE)$ for simplicity.

\smallsection{Degree} The time complexity is $O(\sum_{e\in \SE}|e|)$ if we increment the degree of each of $|e|$ members of each hyperedge $e \in \SE$. 

\smallsection{Coreness} The time complexity is $O(\sum_{e\in \SE}|e|)$, as proven in Theorem 1 of \cite{bu2023hypercore}.

\smallsection{Eigenvector centrality} The time complexity is $O(\sum_{e\in \SE}|e|)$. If we let $I$ be the $|\SV|$ by $|\SE|$ incidence matrix of $\SG$ and $nnz(I)$ be the number of non-zero entries (i.e., $nnz(I)= \sum_{e\in \SE}|e|$), the eigenvector centralities of all nodes in the clique-expanded graph, whose adjacency matrix is $A=II^{T}$, are equivalent to the leading left singular vector of $I$. This can be computed by Power Iteration in $O(nnz(I)T)$ time, where $T$ is the maximum number of iterations. In practice (and in our setting), $T$ is set to a constant, and thus the time complexity is $O(nnz(I))=O(\sum_{e\in \SE}|e|)$, which is even lower than that of materializing the clique-expanded graph.

\smallsection{PageRank} The time complexity is also $O(\sum_{e\in \SE}|e|)$ since its calculation is almost the same as that of the eigenvector centrality. We repeat (at most $T$ times) computing $\mathbf{r}\leftarrow \beta II^T\mathbf{r} + (1-\beta) \mathbf{1}$ for a $|\SV|$-dimensional vector $\mathbf{r}$ and a constant $\beta$. Note that
computing $II^T\mathbf{r}$ can be done in $O(nnz(I))$ time by two steps: (1) computing $\mathbf{r'}\leftarrow I^T\mathbf{r}$ and (2) computing $I\mathbf{r'}$. Since $T$ is set to a constant in practice (and in our setting), the time complexity is $O(nnz(I))=O(\sum_{e\in \SE}|e|)$.


\end{document}